\documentclass[longauth]{aa}

\usepackage{graphicx}
\usepackage{natbib}
\usepackage{scalerel}

\usepackage[table]{xcolor}
\usepackage{orcidlink}
\usepackage{txfonts}
\hypersetup{
    colorlinks=true,
    linkcolor=blue,
    filecolor=magenta,      
    urlcolor=blue,
    citecolor=blue
}
\makeatletter
\renewcommand*\aa@pageof{, page \thepage{} of \pageref*{LastPage}}
\makeatother

\usepackage[utf8]{inputenc}

\usepackage[switch, modulo]{lineno}

\usepackage{euclid}

\newcommand{\euclid}{{\em Euclid}}

\begin{document}
%
%
   \title{\euclid: Early Release Observations -- A preview of the \euclid\ era through a galaxy cluster magnifying lens\thanks{This paper is published on
     behalf of the Euclid Consortium}}

\newcommand{\orcid}[1]{} 
\author{H.~Atek\orcid{0000-0002-7570-0824}\thanks{\email{atek@iap.fr}}\inst{\ref{aff1}}
\and R.~Gavazzi\orcid{0000-0002-5540-6935}\inst{\ref{aff2},\ref{aff1}}
\and J.~R.~Weaver\orcid{0000-0003-1614-196X}\inst{\ref{aff3}}
\and J.~M.~Diego\orcid{0000-0001-9065-3926}\inst{\ref{aff4}}
\and T.~Schrabback\orcid{0000-0002-6987-7834}\inst{\ref{aff5}}
\and N.~A.~Hatch\orcid{0000-0001-5600-0534}\inst{\ref{aff6}}
\and N.~Aghanim\orcid{0000-0002-6688-8992}\inst{\ref{aff7}}
\and H.~Dole\orcid{0000-0002-9767-3839}\inst{\ref{aff7}}
\and W.~G.~Hartley\inst{\ref{aff8}}
\and S.~Taamoli\orcid{0000-0003-0749-4667}\inst{\ref{aff9}}
\and G.~Congedo\orcid{0000-0003-2508-0046}\inst{\ref{aff10}}
\and Y.~Jimenez-Teja\orcid{0000-0002-6090-2853}\inst{\ref{aff11},\ref{aff12}}
\and J.-C.~Cuillandre\orcid{0000-0002-3263-8645}\inst{\ref{aff13}}
\and E.~Ba\~nados\orcid{0000-0002-2931-7824}\inst{\ref{aff14}}
\and S.~Belladitta\orcid{0000-0003-4747-4484}\inst{\ref{aff14},\ref{aff15}}
\and R.~A.~A.~Bowler\orcid{0000-0003-3917-1678}\inst{\ref{aff16}}
\and M.~Franco\orcid{0000-0002-3560-8599}\inst{\ref{aff17}}
\and M.~Jauzac\orcid{0000-0003-1974-8732}\inst{\ref{aff18},\ref{aff19},\ref{aff20},\ref{aff21}}
\and G.~Mahler\orcid{0000-0003-3266-2001}\inst{\ref{aff22},\ref{aff18},\ref{aff19}}
\and J.~Richard\orcid{0000-0001-5492-1049}\inst{\ref{aff23}}
\and P.-F.~Rocci\inst{\ref{aff7}}
\and S.~Serjeant\orcid{0000-0002-0517-7943}\inst{\ref{aff24}}
\and S.~Toft\orcid{0000-0003-3631-7176}\inst{\ref{aff25},\ref{aff26}}
\and D.~Abriola\orcid{0009-0005-4230-3266}\inst{\ref{aff27}}
\and P.~Bergamini\orcid{0000-0003-1383-9414}\inst{\ref{aff27},\ref{aff15}}
\and A.~Biviano\orcid{0000-0002-0857-0732}\inst{\ref{aff28},\ref{aff29}}
\and P.~Dimauro\orcid{0000-0001-7399-2854}\inst{\ref{aff30},\ref{aff12}}
\and M.~Ezziati\orcid{0009-0003-6065-1585}\inst{\ref{aff2}}
\and J.~B.~Golden-Marx\orcid{0000-0002-6394-045X}\inst{\ref{aff6}}
\and C.~Grillo\orcid{0000-0002-5926-7143}\inst{\ref{aff27},\ref{aff31}}
\and A.~C.~N.~Hughes\orcid{0000-0001-9294-3089}\inst{\ref{aff32}}
\and Y.~Kang\orcid{0009-0000-8588-7250}\inst{\ref{aff8}}
\and J.-P.~Kneib\orcid{0000-0002-4616-4989}\inst{\ref{aff33}}
\and M.~Lombardi\orcid{0000-0002-3336-4965}\inst{\ref{aff27}}
\and G.~A.~Mamon\orcid{0000-0001-8956-5953}\inst{\ref{aff1},\ref{aff34}}
\and C.~J.~R.~McPartland\orcid{0000-0003-0639-025X}\inst{\ref{aff35},\ref{aff26}}
\and M.~Meneghetti\orcid{0000-0003-1225-7084}\inst{\ref{aff15},\ref{aff36}}
\and H.~Miyatake\orcid{0000-0001-7964-9766}\inst{\ref{aff37},\ref{aff38},\ref{aff39}}
\and M.~Montes\orcid{0000-0001-7847-0393}\inst{\ref{aff40},\ref{aff41}}
\and D.~J.~Mortlock\orcid{0000-0002-0041-3783}\inst{\ref{aff32},\ref{aff42}}
\and P.~A.~Oesch\orcid{0000-0001-5851-6649}\inst{\ref{aff8},\ref{aff26},\ref{aff25}}
\and N.~Okabe\orcid{0000-0003-2898-0728}\inst{\ref{aff43},\ref{aff44},\ref{aff45}}
\and P.~Rosati\orcid{0000-0002-6813-0632}\inst{\ref{aff46},\ref{aff15}}
\and A.~N.~Taylor\inst{\ref{aff10}}
\and F.~Tarsitano\orcid{0000-0002-5919-0238}\inst{\ref{aff8}}
\and J.~Weller\orcid{0000-0002-8282-2010}\inst{\ref{aff47},\ref{aff48}}
\and M.~Kluge\orcid{0000-0002-9618-2552}\inst{\ref{aff48}}
\and R.~Laureijs\inst{\ref{aff49}}
\and S.~Paltani\orcid{0000-0002-8108-9179}\inst{\ref{aff8}}
\and T.~Saifollahi\orcid{0000-0002-9554-7660}\inst{\ref{aff50},\ref{aff51}}
\and M.~Schirmer\orcid{0000-0003-2568-9994}\inst{\ref{aff14}}
\and C.~Stone\orcid{0000-0002-9086-6398}\inst{\ref{aff52}}
\and A.~Mora\orcid{0000-0002-1922-8529}\inst{\ref{aff53}}
\and B.~Altieri\orcid{0000-0003-3936-0284}\inst{\ref{aff54}}
\and A.~Amara\inst{\ref{aff55}}
\and S.~Andreon\orcid{0000-0002-2041-8784}\inst{\ref{aff56}}
\and N.~Auricchio\orcid{0000-0003-4444-8651}\inst{\ref{aff15}}
\and M.~Baldi\orcid{0000-0003-4145-1943}\inst{\ref{aff57},\ref{aff15},\ref{aff36}}
\and A.~Balestra\orcid{0000-0002-6967-261X}\inst{\ref{aff58}}
\and S.~Bardelli\orcid{0000-0002-8900-0298}\inst{\ref{aff15}}
\and A.~Basset\inst{\ref{aff59}}
\and R.~Bender\orcid{0000-0001-7179-0626}\inst{\ref{aff48},\ref{aff47}}
\and C.~Bodendorf\inst{\ref{aff48}}
\and D.~Bonino\orcid{0000-0002-3336-9977}\inst{\ref{aff60}}
\and E.~Branchini\orcid{0000-0002-0808-6908}\inst{\ref{aff61},\ref{aff62},\ref{aff56}}
\and M.~Brescia\orcid{0000-0001-9506-5680}\inst{\ref{aff63},\ref{aff64},\ref{aff65}}
\and J.~Brinchmann\orcid{0000-0003-4359-8797}\inst{\ref{aff66},\ref{aff67}}
\and S.~Camera\orcid{0000-0003-3399-3574}\inst{\ref{aff68},\ref{aff69},\ref{aff60}}
\and G.~P.~Candini\orcid{0000-0001-9481-8206}\inst{\ref{aff70}}
\and V.~Capobianco\orcid{0000-0002-3309-7692}\inst{\ref{aff60}}
\and C.~Carbone\orcid{0000-0003-0125-3563}\inst{\ref{aff31}}
\and V.~F.~Cardone\inst{\ref{aff30},\ref{aff71}}
\and J.~Carretero\orcid{0000-0002-3130-0204}\inst{\ref{aff72},\ref{aff73}}
\and S.~Casas\orcid{0000-0002-4751-5138}\inst{\ref{aff74}}
\and F.~J.~Castander\orcid{0000-0001-7316-4573}\inst{\ref{aff75},\ref{aff76}}
\and M.~Castellano\orcid{0000-0001-9875-8263}\inst{\ref{aff30}}
\and S.~Cavuoti\orcid{0000-0002-3787-4196}\inst{\ref{aff64},\ref{aff65}}
\and A.~Cimatti\inst{\ref{aff77}}
\and C.~J.~Conselice\orcid{0000-0003-1949-7638}\inst{\ref{aff16}}
\and L.~Conversi\orcid{0000-0002-6710-8476}\inst{\ref{aff78},\ref{aff54}}
\and Y.~Copin\orcid{0000-0002-5317-7518}\inst{\ref{aff79}}
\and L.~Corcione\orcid{0000-0002-6497-5881}\inst{\ref{aff60}}
\and F.~Courbin\orcid{0000-0003-0758-6510}\inst{\ref{aff33}}
\and H.~M.~Courtois\orcid{0000-0003-0509-1776}\inst{\ref{aff80}}
\and A.~Da~Silva\orcid{0000-0002-6385-1609}\inst{\ref{aff81},\ref{aff82}}
\and H.~Degaudenzi\orcid{0000-0002-5887-6799}\inst{\ref{aff8}}
\and A.~M.~Di~Giorgio\orcid{0000-0002-4767-2360}\inst{\ref{aff83}}
\and J.~Dinis\orcid{0000-0001-5075-1601}\inst{\ref{aff81},\ref{aff82}}
\and M.~Douspis\orcid{0000-0003-4203-3954}\inst{\ref{aff7}}
\and F.~Dubath\orcid{0000-0002-6533-2810}\inst{\ref{aff8}}
\and X.~Dupac\inst{\ref{aff54}}
\and S.~Dusini\orcid{0000-0002-1128-0664}\inst{\ref{aff84}}
\and A.~Ealet\orcid{0000-0003-3070-014X}\inst{\ref{aff79}}
\and M.~Farina\orcid{0000-0002-3089-7846}\inst{\ref{aff83}}
\and S.~Farrens\orcid{0000-0002-9594-9387}\inst{\ref{aff13}}
\and S.~Ferriol\inst{\ref{aff79}}
\and S.~Fotopoulou\orcid{0000-0002-9686-254X}\inst{\ref{aff85}}
\and M.~Frailis\orcid{0000-0002-7400-2135}\inst{\ref{aff28}}
\and E.~Franceschi\orcid{0000-0002-0585-6591}\inst{\ref{aff15}}
\and S.~Galeotta\orcid{0000-0002-3748-5115}\inst{\ref{aff28}}
\and W.~Gillard\orcid{0000-0003-4744-9748}\inst{\ref{aff86}}
\and B.~Gillis\orcid{0000-0002-4478-1270}\inst{\ref{aff10}}
\and C.~Giocoli\orcid{0000-0002-9590-7961}\inst{\ref{aff15},\ref{aff87}}
\and P.~G\'omez-Alvarez\orcid{0000-0002-8594-5358}\inst{\ref{aff88},\ref{aff54}}
\and A.~Grazian\orcid{0000-0002-5688-0663}\inst{\ref{aff58}}
\and F.~Grupp\inst{\ref{aff48},\ref{aff47}}
\and L.~Guzzo\orcid{0000-0001-8264-5192}\inst{\ref{aff27},\ref{aff56}}
\and M.~Hailey\inst{\ref{aff70}}
\and S.~V.~H.~Haugan\orcid{0000-0001-9648-7260}\inst{\ref{aff89}}
\and J.~Hoar\inst{\ref{aff54}}
\and H.~Hoekstra\orcid{0000-0002-0641-3231}\inst{\ref{aff90}}
\and M.~S.~Holliman\inst{\ref{aff10}}
\and W.~Holmes\inst{\ref{aff91}}
\and I.~Hook\orcid{0000-0002-2960-978X}\inst{\ref{aff92}}
\and F.~Hormuth\inst{\ref{aff93}}
\and A.~Hornstrup\orcid{0000-0002-3363-0936}\inst{\ref{aff94},\ref{aff35}}
\and P.~Hudelot\inst{\ref{aff1}}
\and K.~Jahnke\orcid{0000-0003-3804-2137}\inst{\ref{aff14}}
\and M.~Jhabvala\inst{\ref{aff95}}
\and E.~Keih\"anen\orcid{0000-0003-1804-7715}\inst{\ref{aff96}}
\and S.~Kermiche\orcid{0000-0002-0302-5735}\inst{\ref{aff86}}
\and A.~Kiessling\orcid{0000-0002-2590-1273}\inst{\ref{aff91}}
\and T.~Kitching\orcid{0000-0002-4061-4598}\inst{\ref{aff70}}
\and R.~Kohley\inst{\ref{aff54}}
\and B.~Kubik\orcid{0009-0006-5823-4880}\inst{\ref{aff79}}
\and K.~Kuijken\orcid{0000-0002-3827-0175}\inst{\ref{aff90}}
\and M.~K\"ummel\orcid{0000-0003-2791-2117}\inst{\ref{aff47}}
\and M.~Kunz\orcid{0000-0002-3052-7394}\inst{\ref{aff97}}
\and H.~Kurki-Suonio\orcid{0000-0002-4618-3063}\inst{\ref{aff98},\ref{aff99}}
\and O.~Lahav\orcid{0000-0002-1134-9035}\inst{\ref{aff100}}
\and D.~Le~Mignant\orcid{0000-0002-5339-5515}\inst{\ref{aff2}}
\and S.~Ligori\orcid{0000-0003-4172-4606}\inst{\ref{aff60}}
\and P.~B.~Lilje\orcid{0000-0003-4324-7794}\inst{\ref{aff89}}
\and V.~Lindholm\orcid{0000-0003-2317-5471}\inst{\ref{aff98},\ref{aff99}}
\and I.~Lloro\inst{\ref{aff101}}
\and D.~Maino\inst{\ref{aff27},\ref{aff31},\ref{aff102}}
\and E.~Maiorano\orcid{0000-0003-2593-4355}\inst{\ref{aff15}}
\and O.~Mansutti\orcid{0000-0001-5758-4658}\inst{\ref{aff28}}
\and O.~Marggraf\orcid{0000-0001-7242-3852}\inst{\ref{aff103}}
\and K.~Markovic\orcid{0000-0001-6764-073X}\inst{\ref{aff91}}
\and N.~Martinet\orcid{0000-0003-2786-7790}\inst{\ref{aff2}}
\and F.~Marulli\orcid{0000-0002-8850-0303}\inst{\ref{aff104},\ref{aff15},\ref{aff36}}
\and R.~Massey\orcid{0000-0002-6085-3780}\inst{\ref{aff18},\ref{aff19}}
\and S.~Maurogordato\inst{\ref{aff105}}
\and H.~J.~McCracken\orcid{0000-0002-9489-7765}\inst{\ref{aff1}}
\and S.~Mei\orcid{0000-0002-2849-559X}\inst{\ref{aff106}}
\and Y.~Mellier\inst{\ref{aff34},\ref{aff1}}
\and E.~Merlin\orcid{0000-0001-6870-8900}\inst{\ref{aff30}}
\and G.~Meylan\inst{\ref{aff33}}
\and M.~Moresco\orcid{0000-0002-7616-7136}\inst{\ref{aff104},\ref{aff15}}
\and L.~Moscardini\orcid{0000-0002-3473-6716}\inst{\ref{aff104},\ref{aff15},\ref{aff36}}
\and R.~Nakajima\inst{\ref{aff103}}
\and R.~C.~Nichol\orcid{0000-0003-0939-6518}\inst{\ref{aff55}}
\and S.-M.~Niemi\inst{\ref{aff49}}
\and C.~Padilla\orcid{0000-0001-7951-0166}\inst{\ref{aff107}}
\and K.~Paech\orcid{0000-0003-0625-2367}\inst{\ref{aff48}}
\and F.~Pasian\orcid{0000-0002-4869-3227}\inst{\ref{aff28}}
\and J.~A.~Peacock\orcid{0000-0002-1168-8299}\inst{\ref{aff10}}
\and K.~Pedersen\inst{\ref{aff108}}
\and W.~J.~Percival\orcid{0000-0002-0644-5727}\inst{\ref{aff109},\ref{aff110},\ref{aff111}}
\and V.~Pettorino\inst{\ref{aff49}}
\and S.~Pires\orcid{0000-0002-0249-2104}\inst{\ref{aff13}}
\and G.~Polenta\orcid{0000-0003-4067-9196}\inst{\ref{aff112}}
\and M.~Poncet\inst{\ref{aff59}}
\and L.~A.~Popa\inst{\ref{aff113}}
\and L.~Pozzetti\orcid{0000-0001-7085-0412}\inst{\ref{aff15}}
\and F.~Raison\orcid{0000-0002-7819-6918}\inst{\ref{aff48}}
\and A.~Renzi\orcid{0000-0001-9856-1970}\inst{\ref{aff114},\ref{aff84}}
\and J.~Rhodes\orcid{0000-0002-4485-8549}\inst{\ref{aff91}}
\and G.~Riccio\inst{\ref{aff64}}
\and E.~Romelli\orcid{0000-0003-3069-9222}\inst{\ref{aff28}}
\and M.~Roncarelli\orcid{0000-0001-9587-7822}\inst{\ref{aff15}}
\and R.~Saglia\orcid{0000-0003-0378-7032}\inst{\ref{aff47},\ref{aff48}}
\and D.~Sapone\orcid{0000-0001-7089-4503}\inst{\ref{aff115}}
\and P.~Schneider\orcid{0000-0001-8561-2679}\inst{\ref{aff103}}
\and A.~Secroun\orcid{0000-0003-0505-3710}\inst{\ref{aff86}}
\and G.~Seidel\orcid{0000-0003-2907-353X}\inst{\ref{aff14}}
\and S.~Serrano\orcid{0000-0002-0211-2861}\inst{\ref{aff76},\ref{aff116},\ref{aff75}}
\and C.~Sirignano\orcid{0000-0002-0995-7146}\inst{\ref{aff114},\ref{aff84}}
\and G.~Sirri\orcid{0000-0003-2626-2853}\inst{\ref{aff36}}
\and J.~Skottfelt\orcid{0000-0003-1310-8283}\inst{\ref{aff117}}
\and L.~Stanco\orcid{0000-0002-9706-5104}\inst{\ref{aff84}}
\and P.~Tallada-Cresp\'{i}\orcid{0000-0002-1336-8328}\inst{\ref{aff72},\ref{aff73}}
\and H.~I.~Teplitz\orcid{0000-0002-7064-5424}\inst{\ref{aff118}}
\and I.~Tereno\inst{\ref{aff81},\ref{aff119}}
\and R.~Toledo-Moreo\orcid{0000-0002-2997-4859}\inst{\ref{aff120}}
\and I.~Tutusaus\orcid{0000-0002-3199-0399}\inst{\ref{aff121}}
\and L.~Valenziano\orcid{0000-0002-1170-0104}\inst{\ref{aff15},\ref{aff122}}
\and T.~Vassallo\orcid{0000-0001-6512-6358}\inst{\ref{aff47},\ref{aff28}}
\and G.~Verdoes~Kleijn\orcid{0000-0001-5803-2580}\inst{\ref{aff51}}
\and A.~Veropalumbo\orcid{0000-0003-2387-1194}\inst{\ref{aff56},\ref{aff62},\ref{aff123}}
\and Y.~Wang\orcid{0000-0002-4749-2984}\inst{\ref{aff118}}
\and E.~Zucca\orcid{0000-0002-5845-8132}\inst{\ref{aff15}}
\and C.~Baccigalupi\orcid{0000-0002-8211-1630}\inst{\ref{aff29},\ref{aff28},\ref{aff124},\ref{aff125}}
\and C.~Burigana\orcid{0000-0002-3005-5796}\inst{\ref{aff126},\ref{aff122}}
\and G.~Castignani\orcid{0000-0001-6831-0687}\inst{\ref{aff15}}
\and Z.~Sakr\orcid{0000-0002-4823-3757}\inst{\ref{aff127},\ref{aff121},\ref{aff128}}
\and V.~Scottez\inst{\ref{aff34},\ref{aff129}}
\and M.~Viel\orcid{0000-0002-2642-5707}\inst{\ref{aff29},\ref{aff28},\ref{aff125},\ref{aff124},\ref{aff130}}
\and P.~Simon\inst{\ref{aff103}}
\and D.~Stern\orcid{0000-0003-2686-9241}\inst{\ref{aff91}}
\and J.~Mart\'{i}n-Fleitas\orcid{0000-0002-8594-569X}\inst{\ref{aff53}}
\and D.~Scott\orcid{0000-0002-6878-9840}\inst{\ref{aff131}}}
										   
\institute{Institut d'Astrophysique de Paris, UMR 7095, CNRS, and Sorbonne Universit\'e, 98 bis boulevard Arago, 75014 Paris, France\label{aff1}
\and
Aix-Marseille Universit\'e, CNRS, CNES, LAM, Marseille, France\label{aff2}
\and
Department of Astronomy, University of Massachusetts, Amherst, MA 01003, USA\label{aff3}
\and
Instituto de F\'isica de Cantabria, Edificio Juan Jord\'a, Avenida de los Castros, 39005 Santander, Spain\label{aff4}
\and
Universit\"at Innsbruck, Institut f\"ur Astro- und Teilchenphysik, Technikerstr. 25/8, 6020 Innsbruck, Austria\label{aff5}
\and
School of Physics and Astronomy, University of Nottingham, University Park, Nottingham NG7 2RD, UK\label{aff6}
\and
Universit\'e Paris-Saclay, CNRS, Institut d'astrophysique spatiale, 91405, Orsay, France\label{aff7}
\and
Department of Astronomy, University of Geneva, ch. d'Ecogia 16, 1290 Versoix, Switzerland\label{aff8}
\and
Physics and Astronomy Department, University of California, 900 University Ave., Riverside, CA 92521, USA\label{aff9}
\and
Institute for Astronomy, University of Edinburgh, Royal Observatory, Blackford Hill, Edinburgh EH9 3HJ, UK\label{aff10}
\and
Instituto de Astrof\'isica de Andaluc\'ia, CSIC, Glorieta de la Astronom\'\i a, 18080, Granada, Spain\label{aff11}
\and
Observatorio Nacional, Rua General Jose Cristino, 77-Bairro Imperial de Sao Cristovao, Rio de Janeiro, 20921-400, Brazil\label{aff12}
\and
Universit\'e Paris-Saclay, Universit\'e Paris Cit\'e, CEA, CNRS, AIM, 91191, Gif-sur-Yvette, France\label{aff13}
\and
Max-Planck-Institut f\"ur Astronomie, K\"onigstuhl 17, 69117 Heidelberg, Germany\label{aff14}
\and
INAF-Osservatorio di Astrofisica e Scienza dello Spazio di Bologna, Via Piero Gobetti 93/3, 40129 Bologna, Italy\label{aff15}
\and
Jodrell Bank Centre for Astrophysics, Department of Physics and Astronomy, University of Manchester, Oxford Road, Manchester M13 9PL, UK\label{aff16}
\and
The University of Texas at Austin, Austin, TX, 78712, USA\label{aff17}
\and
Department of Physics, Centre for Extragalactic Astronomy, Durham University, South Road, DH1 3LE, UK\label{aff18}
\and
Department of Physics, Institute for Computational Cosmology, Durham University, South Road, DH1 3LE, UK\label{aff19}
\and
Astrophysics Research Centre, University of KwaZulu-Natal, Westville Campus, Durban 4041, South Africa\label{aff20}
\and
School of Mathematics, Statistics \& Computer Science, University of KwaZulu-Natal, Westville Campus, Durban 4041, South Africa\label{aff21}
\and
STAR Institute, Quartier Agora - All\'ee du six Ao\^ut, 19c B-4000 Li\`ege, Belgium\label{aff22}
\and
Centre de Recherche Astrophysique de Lyon, UMR5574, CNRS, Universit\'e Claude Bernard Lyon 1, ENS de Lyon, 69230, Saint-Genis-Laval, France\label{aff23}
\and
School of Physical Sciences, The Open University, Milton Keynes, MK7 6AA, UK\label{aff24}
\and
Cosmic Dawn Center (DAWN)\label{aff25}
\and
Niels Bohr Institute, University of Copenhagen, Jagtvej 128, 2200 Copenhagen, Denmark\label{aff26}
\and
Dipartimento di Fisica "Aldo Pontremoli", Universit\`a degli Studi di Milano, Via Celoria 16, 20133 Milano, Italy\label{aff27}
\and
INAF-Osservatorio Astronomico di Trieste, Via G. B. Tiepolo 11, 34143 Trieste, Italy\label{aff28}
\and
IFPU, Institute for Fundamental Physics of the Universe, via Beirut 2, 34151 Trieste, Italy\label{aff29}
\and
INAF-Osservatorio Astronomico di Roma, Via Frascati 33, 00078 Monteporzio Catone, Italy\label{aff30}
\and
INAF-IASF Milano, Via Alfonso Corti 12, 20133 Milano, Italy\label{aff31}
\and
Astrophysics Group, Blackett Laboratory, Imperial College London, London SW7 2AZ, UK\label{aff32}
\and
Institute of Physics, Laboratory of Astrophysics, Ecole Polytechnique F\'ed\'erale de Lausanne (EPFL), Observatoire de Sauverny, 1290 Versoix, Switzerland\label{aff33}
\and
Institut d'Astrophysique de Paris, 98bis Boulevard Arago, 75014, Paris, France\label{aff34}
\and
Cosmic Dawn Center (DAWN), Denmark\label{aff35}
\and
INFN-Sezione di Bologna, Viale Berti Pichat 6/2, 40127 Bologna, Italy\label{aff36}
\and
Kobayashi-Maskawa Institute for the Origin of Particles and the Universe, Nagoya University, Chikusa-ku, Nagoya, 464-8602, Japan\label{aff37}
\and
Institute for Advanced Research, Nagoya University, Chikusa-ku, Nagoya, 464-8601, Japan\label{aff38}
\and
Kavli Institute for the Physics and Mathematics of the Universe (WPI), University of Tokyo, Kashiwa, Chiba 277-8583, Japan\label{aff39}
\and
Instituto de Astrof\'isica de Canarias, Calle V\'ia L\'actea s/n, 38204, San Crist\'obal de La Laguna, Tenerife, Spain\label{aff40}
\and
Departamento de Astrof\'isica, Universidad de La Laguna, 38206, La Laguna, Tenerife, Spain\label{aff41}
\and
Department of Mathematics, Imperial College London, London SW7 2AZ, UK\label{aff42}
\and
Physics Program, Graduate School of Advanced Science and Engineering, Hiroshima University, 1-3-1 Kagamiyama, Higashi-Hiroshima, Hiroshima 739-8526, Japan\label{aff43}
\and
Hiroshima Astrophysical Science Center, Hiroshima University, 1-3-1 Kagamiyama, Higashi-Hiroshima, Hiroshima 739-8526, Japan\label{aff44}
\and
Core Research for Energetic Universe, Hiroshima University, 1-3-1, Kagamiyama, Higashi-Hiroshima, Hiroshima 739-8526, Japan\label{aff45}
\and
Dipartimento di Fisica e Scienze della Terra, Universit\`a degli Studi di Ferrara, Via Giuseppe Saragat 1, 44122 Ferrara, Italy\label{aff46}
\and
Universit\"ats-Sternwarte M\"unchen, Fakult\"at f\"ur Physik, Ludwig-Maximilians-Universit\"at M\"unchen, Scheinerstrasse 1, 81679 M\"unchen, Germany\label{aff47}
\and
Max Planck Institute for Extraterrestrial Physics, Giessenbachstr. 1, 85748 Garching, Germany\label{aff48}
\and
European Space Agency/ESTEC, Keplerlaan 1, 2201 AZ Noordwijk, The Netherlands\label{aff49}
\and
Observatoire Astronomique de Strasbourg (ObAS), Universit\'e de Strasbourg - CNRS, UMR 7550, Strasbourg, France\label{aff50}
\and
Kapteyn Astronomical Institute, University of Groningen, PO Box 800, 9700 AV Groningen, The Netherlands\label{aff51}
\and
Department of Physics, Universit\'{e} de Montr\'{e}al, 2900 Edouard Montpetit Blvd, Montr\'{e}al, Qu\'{e}bec H3T 1J4, Canada\label{aff52}
\and
Aurora Technology for European Space Agency (ESA), Camino bajo del Castillo, s/n, Urbanizacion Villafranca del Castillo, Villanueva de la Ca\~nada, 28692 Madrid, Spain\label{aff53}
\and
ESAC/ESA, Camino Bajo del Castillo, s/n., Urb. Villafranca del Castillo, 28692 Villanueva de la Ca\~nada, Madrid, Spain\label{aff54}
\and
School of Mathematics and Physics, University of Surrey, Guildford, Surrey, GU2 7XH, UK\label{aff55}
\and
INAF-Osservatorio Astronomico di Brera, Via Brera 28, 20122 Milano, Italy\label{aff56}
\and
Dipartimento di Fisica e Astronomia, Universit\`a di Bologna, Via Gobetti 93/2, 40129 Bologna, Italy\label{aff57}
\and
INAF-Osservatorio Astronomico di Padova, Via dell'Osservatorio 5, 35122 Padova, Italy\label{aff58}
\and
Centre National d'Etudes Spatiales -- Centre spatial de Toulouse, 18 avenue Edouard Belin, 31401 Toulouse Cedex 9, France\label{aff59}
\and
INAF-Osservatorio Astrofisico di Torino, Via Osservatorio 20, 10025 Pino Torinese (TO), Italy\label{aff60}
\and
Dipartimento di Fisica, Universit\`a di Genova, Via Dodecaneso 33, 16146, Genova, Italy\label{aff61}
\and
INFN-Sezione di Genova, Via Dodecaneso 33, 16146, Genova, Italy\label{aff62}
\and
Department of Physics "E. Pancini", University Federico II, Via Cinthia 6, 80126, Napoli, Italy\label{aff63}
\and
INAF-Osservatorio Astronomico di Capodimonte, Via Moiariello 16, 80131 Napoli, Italy\label{aff64}
\and
INFN section of Naples, Via Cinthia 6, 80126, Napoli, Italy\label{aff65}
\and
Instituto de Astrof\'isica e Ci\^encias do Espa\c{c}o, Universidade do Porto, CAUP, Rua das Estrelas, PT4150-762 Porto, Portugal\label{aff66}
\and
Faculdade de Ci\^encias da Universidade do Porto, Rua do Campo de Alegre, 4150-007 Porto, Portugal\label{aff67}
\and
Dipartimento di Fisica, Universit\`a degli Studi di Torino, Via P. Giuria 1, 10125 Torino, Italy\label{aff68}
\and
INFN-Sezione di Torino, Via P. Giuria 1, 10125 Torino, Italy\label{aff69}
\and
Mullard Space Science Laboratory, University College London, Holmbury St Mary, Dorking, Surrey RH5 6NT, UK\label{aff70}
\and
INFN-Sezione di Roma, Piazzale Aldo Moro, 2 - c/o Dipartimento di Fisica, Edificio G. Marconi, 00185 Roma, Italy\label{aff71}
\and
Centro de Investigaciones Energ\'eticas, Medioambientales y Tecnol\'ogicas (CIEMAT), Avenida Complutense 40, 28040 Madrid, Spain\label{aff72}
\and
Port d'Informaci\'{o} Cient\'{i}fica, Campus UAB, C. Albareda s/n, 08193 Bellaterra (Barcelona), Spain\label{aff73}
\and
Institute for Theoretical Particle Physics and Cosmology (TTK), RWTH Aachen University, 52056 Aachen, Germany\label{aff74}
\and
Institute of Space Sciences (ICE, CSIC), Campus UAB, Carrer de Can Magrans, s/n, 08193 Barcelona, Spain\label{aff75}
\and
Institut d'Estudis Espacials de Catalunya (IEEC),  Edifici RDIT, Campus UPC, 08860 Castelldefels, Barcelona, Spain\label{aff76}
\and
Dipartimento di Fisica e Astronomia "Augusto Righi" - Alma Mater Studiorum Universit\`a di Bologna, Viale Berti Pichat 6/2, 40127 Bologna, Italy\label{aff77}
\and
European Space Agency/ESRIN, Largo Galileo Galilei 1, 00044 Frascati, Roma, Italy\label{aff78}
\and
Universit\'e Claude Bernard Lyon 1, CNRS/IN2P3, IP2I Lyon, UMR 5822, Villeurbanne, F-69100, France\label{aff79}
\and
UCB Lyon 1, CNRS/IN2P3, IUF, IP2I Lyon, 4 rue Enrico Fermi, 69622 Villeurbanne, France\label{aff80}
\and
Departamento de F\'isica, Faculdade de Ci\^encias, Universidade de Lisboa, Edif\'icio C8, Campo Grande, PT1749-016 Lisboa, Portugal\label{aff81}
\and
Instituto de Astrof\'isica e Ci\^encias do Espa\c{c}o, Faculdade de Ci\^encias, Universidade de Lisboa, Campo Grande, 1749-016 Lisboa, Portugal\label{aff82}
\and
INAF-Istituto di Astrofisica e Planetologia Spaziali, via del Fosso del Cavaliere, 100, 00100 Roma, Italy\label{aff83}
\and
INFN-Padova, Via Marzolo 8, 35131 Padova, Italy\label{aff84}
\and
School of Physics, HH Wills Physics Laboratory, University of Bristol, Tyndall Avenue, Bristol, BS8 1TL, UK\label{aff85}
\and
Aix-Marseille Universit\'e, CNRS/IN2P3, CPPM, Marseille, France\label{aff86}
\and
Istituto Nazionale di Fisica Nucleare, Sezione di Bologna, Via Irnerio 46, 40126 Bologna, Italy\label{aff87}
\and
FRACTAL S.L.N.E., calle Tulip\'an 2, Portal 13 1A, 28231, Las Rozas de Madrid, Spain\label{aff88}
\and
Institute of Theoretical Astrophysics, University of Oslo, P.O. Box 1029 Blindern, 0315 Oslo, Norway\label{aff89}
\and
Leiden Observatory, Leiden University, Einsteinweg 55, 2333 CC Leiden, The Netherlands\label{aff90}
\and
Jet Propulsion Laboratory, California Institute of Technology, 4800 Oak Grove Drive, Pasadena, CA, 91109, USA\label{aff91}
\and
Department of Physics, Lancaster University, Lancaster, LA1 4YB, UK\label{aff92}
\and
Felix Hormuth Engineering, Goethestr. 17, 69181 Leimen, Germany\label{aff93}
\and
Technical University of Denmark, Elektrovej 327, 2800 Kgs. Lyngby, Denmark\label{aff94}
\and
NASA Goddard Space Flight Center, Greenbelt, MD 20771, USA\label{aff95}
\and
Department of Physics and Helsinki Institute of Physics, Gustaf H\"allstr\"omin katu 2, 00014 University of Helsinki, Finland\label{aff96}
\and
Universit\'e de Gen\`eve, D\'epartement de Physique Th\'eorique and Centre for Astroparticle Physics, 24 quai Ernest-Ansermet, CH-1211 Gen\`eve 4, Switzerland\label{aff97}
\and
Department of Physics, P.O. Box 64, 00014 University of Helsinki, Finland\label{aff98}
\and
Helsinki Institute of Physics, Gustaf H{\"a}llstr{\"o}min katu 2, University of Helsinki, Helsinki, Finland\label{aff99}
\and
Department of Physics and Astronomy, University College London, Gower Street, London WC1E 6BT, UK\label{aff100}
\and
NOVA optical infrared instrumentation group at ASTRON, Oude Hoogeveensedijk 4, 7991PD, Dwingeloo, The Netherlands\label{aff101}
\and
INFN-Sezione di Milano, Via Celoria 16, 20133 Milano, Italy\label{aff102}
\and
Universit\"at Bonn, Argelander-Institut f\"ur Astronomie, Auf dem H\"ugel 71, 53121 Bonn, Germany\label{aff103}
\and
Dipartimento di Fisica e Astronomia "Augusto Righi" - Alma Mater Studiorum Universit\`a di Bologna, via Piero Gobetti 93/2, 40129 Bologna, Italy\label{aff104}
\and
Universit\'e C\^{o}te d'Azur, Observatoire de la C\^{o}te d'Azur, CNRS, Laboratoire Lagrange, Bd de l'Observatoire, CS 34229, 06304 Nice cedex 4, France\label{aff105}
\and
Universit\'e Paris Cit\'e, CNRS, Astroparticule et Cosmologie, 75013 Paris, France\label{aff106}
\and
Institut de F\'{i}sica d'Altes Energies (IFAE), The Barcelona Institute of Science and Technology, Campus UAB, 08193 Bellaterra (Barcelona), Spain\label{aff107}
\and
Department of Physics and Astronomy, University of Aarhus, Ny Munkegade 120, DK-8000 Aarhus C, Denmark\label{aff108}
\and
Waterloo Centre for Astrophysics, University of Waterloo, Waterloo, Ontario N2L 3G1, Canada\label{aff109}
\and
Department of Physics and Astronomy, University of Waterloo, Waterloo, Ontario N2L 3G1, Canada\label{aff110}
\and
Perimeter Institute for Theoretical Physics, Waterloo, Ontario N2L 2Y5, Canada\label{aff111}
\and
Space Science Data Center, Italian Space Agency, via del Politecnico snc, 00133 Roma, Italy\label{aff112}
\and
Institute of Space Science, Str. Atomistilor, nr. 409 M\u{a}gurele, Ilfov, 077125, Romania\label{aff113}
\and
Dipartimento di Fisica e Astronomia "G. Galilei", Universit\`a di Padova, Via Marzolo 8, 35131 Padova, Italy\label{aff114}
\and
Departamento de F\'isica, FCFM, Universidad de Chile, Blanco Encalada 2008, Santiago, Chile\label{aff115}
\and
Satlantis, University Science Park, Sede Bld 48940, Leioa-Bilbao, Spain\label{aff116}
\and
Centre for Electronic Imaging, Open University, Walton Hall, Milton Keynes, MK7~6AA, UK\label{aff117}
\and
Infrared Processing and Analysis Center, California Institute of Technology, Pasadena, CA 91125, USA\label{aff118}
\and
Instituto de Astrof\'isica e Ci\^encias do Espa\c{c}o, Faculdade de Ci\^encias, Universidade de Lisboa, Tapada da Ajuda, 1349-018 Lisboa, Portugal\label{aff119}
\and
Universidad Polit\'ecnica de Cartagena, Departamento de Electr\'onica y Tecnolog\'ia de Computadoras,  Plaza del Hospital 1, 30202 Cartagena, Spain\label{aff120}
\and
Institut de Recherche en Astrophysique et Plan\'etologie (IRAP), Universit\'e de Toulouse, CNRS, UPS, CNES, 14 Av. Edouard Belin, 31400 Toulouse, France\label{aff121}
\and
INFN-Bologna, Via Irnerio 46, 40126 Bologna, Italy\label{aff122}
\and
Dipartimento di Fisica, Universit\`a degli studi di Genova, and INFN-Sezione di Genova, via Dodecaneso 33, 16146, Genova, Italy\label{aff123}
\and
INFN, Sezione di Trieste, Via Valerio 2, 34127 Trieste TS, Italy\label{aff124}
\and
SISSA, International School for Advanced Studies, Via Bonomea 265, 34136 Trieste TS, Italy\label{aff125}
\and
INAF, Istituto di Radioastronomia, Via Piero Gobetti 101, 40129 Bologna, Italy\label{aff126}
\and
Institut f\"ur Theoretische Physik, University of Heidelberg, Philosophenweg 16, 69120 Heidelberg, Germany\label{aff127}
\and
Universit\'e St Joseph; Faculty of Sciences, Beirut, Lebanon\label{aff128}
\and
Junia, EPA department, 41 Bd Vauban, 59800 Lille, France\label{aff129}
\and
ICSC - Centro Nazionale di Ricerca in High Performance Computing, Big Data e Quantum Computing, Via Magnanelli 2, Bologna, Italy\label{aff130}
\and
Department of Physics and Astronomy, University of British Columbia, Vancouver, BC V6T 1Z1, Canada\label{aff131}}    
%
\abstract{We present the first analysis of the \euclid\ Early Release Observations (ERO) program that targets fields around two lensing clusters, Abell 2390 and Abell 2764. We use VIS and NISP imaging to produce photometric catalogs for a total of $\sim 500\,000$ objects. The imaging data reach a $5\,\sigma$ typical depth in the range 25.1--25.4 AB in the NISP bands, and 27.1--27.3 AB in the VIS band. Using the Lyman-break method in combination with photometric redshifts, we search for high-redshift galaxies. We identify $30$ Lyman-break galaxy (LBG) candidates at $z>6$ and 139 extremely red sources (ERSs), most likely at lower redshift. The deeper VIS imaging compared to NISP means we can routinely identify high-redshift Lyman breaks of the order of $3$ magnitudes, which reduces contamination by brown dwarf stars and low-redshift galaxies. The difficulty of spatially resolving most of these sources in 0\farcs3 pix$^{-1}$ imaging makes the distinction between galaxies and quasars more challenging. Spectroscopic follow-up campaigns of such bright sources will help constrain both the bright end of the ultraviolet galaxy luminosity function and the quasar luminosity function at $z>6$, and constrain the physical nature of these objects. Additionally, we have performed a combined strong lensing and weak lensing analysis of A2390, and demonstrate how \euclid\ will contribute to better constraining the virial mass of galaxy clusters. From these data, we also identify optical and near-infrared counterparts of known $z>0.6$ clusters, which exhibit strong lensing features, establishing the ability of \euclid\ to characterize high-redshift clusters. Finally, we provide a glimpse of \euclid's ability to map the intracluster light out to larger radii than current facilities, enabling a better understanding of the cluster assembly history and mapping of the dark matter distribution. This initial dataset illustrates the diverse spectrum of legacy science that will be enabled by the \euclid\ survey. 
}
%
%
%
   \titlerunning{\Euclid: ERO -- Magnifying lens}
   \authorrunning{Atek et al.}
   
   \maketitle
%
%
%
%
   
\section{\label{sc:Intro}Introduction}

Beyond its primary cosmology science goals, the \euclid\ mission will provide a sample of more than 200\,000 high-redshift galaxies across 14\,000\,deg$^2$ of extragalactic sky \citep{Scaramella-EP1, EuclidSkyOverview}. This will enable a wide array of legacy science. In particular, the near-infrared (NIR) capabilities of the NISP (Near-Infrared
Spectrometer and Photometer) instrument \citep{EuclidSkyNISP} and the wide optical coverage of the visible instrument \citep[VIS;][]{EuclidSkyVIS}, allow us to detect star-forming galaxies and quasars during the epoch of reionization, ranging from $z=6$ to $z=9$.

The Deep Survey component of \euclid\ will map a total of 53 deg$^2$, namely in the Euclid Deep Field North (EDF-N), the Euclid Deep Field South (EDF-S), and the Euclid Deep Field Fornax (EDF-F), down to a $5\,\sigma$ limiting magnitude of $\sim 26.2$ in the NIR and 28.5 in the visible. In comparison, the widest areas covered by the \HST (HST) and the JWST, are of the order of 1 deg$^{2}$. The Wide Survey will reach a $5\,\sigma$ magnitude depth of 26.2 in VIS/$\IE$ and 24.5 in all NISP bands. The sky coverage is optimized to have maximum overlap with ground-based ancillary data, which greatly help with photometric redshift estimates \citep{vanMierlo-EP21,Scaramella-EP1,EuclidSkyOverview}.  

One of the main legacy science components is the identification of distant galaxies and quasars at redshifts greater than 6. This will rely primarily on the detection of Lyman break in the VIS $\IE$ filter at $z\sim7$, and in one of the NISP filters at higher redshifts. Based on the current survey design,  more than 10\,000 galaxies are expected at $z \sim 7$ and up to 2000 galaxies at $z\sim 8$, assuming a Schechter UV luminosity function \citep{bouwens15}. Over 100 quasars are expected over $7.0<z<7.5$, and about 25 quasars beyond $z=7.5$, assuming a declining quasar luminosity function beyond $z=6$ \citep{Barnett-EP5}. The brightness of these sources, \JE$\gtrsim 21$\, opens up spectroscopic follow-up opportunities with ground-based telescopes to robustly determine the bright-end of the galaxy and the quasar luminosity functions. In addition, \euclid\ will observe hundreds of lensing clusters, enabling the detection of magnified galaxies and quasars at $z>6$.

Here we present an overview and early results of the Early Release Observations (ERO) program Magnifying Lens, which targets two cluster fields, Abell 2390 and Abell 2764 (hereafter A2390 and A2764, respectively). This program represents a unique opportunity to showcase the science questions that \Euclid's observations of galaxy clusters will be able to address. This paper serves as an introduction to these initial measurements and findings in both cluster and blank fields. Specifically, it illustrates how \Euclid will identify high-redshift dropout sources in both blank regions and behind lensing clusters. A companion paper is dedicated to the search for high-redshift candidates at $z>6$ \citep[see][]{EROLensVISdropouts}. We also take this opportunity to investigate the challenges in robustly identifying high-redshift objects and the potential sources of contamination. These observations also offer a glimpse into the combination of strong lensing and weak lensing analyses to map the mass distribution in lensing clusters that will be uncovered in \euclid\, data. Furthermore, \euclid\ offers a wide field of view that is necessary to map the intracluster light in the outer regions of the cluster.   

The paper is organized as follows. In Sect. \ref{sec:ero}, we describe the ERO program and the characteristics of the targets. Section \ref{sec:obs} is dedicated to the observations and data reduction of the VIS and NISP imaging. The photometric catalogs that are released with this publication are described in Sect. \ref{sec:catalogs}. In terms of scientific exploitation of this dataset, we present the VIS-dropout sources in Sect. \ref{sec:dropouts}, the weak lensing analysis of A2390 in Sect. \ref{sec:wl} and a combined strong lensing and weak lensing analysis of A2390 in Sect. \ref{sec:sl}. The search for high-redshift galaxy clusters is discussed in Sect. \ref{sec:erosita}, while the study of the intracluster light in A2390 is presented in Sect. \ref{sec:icl}. A summary of this work is presented in Sect. \ref{sec:summary}. We assume a flat $\Lambda$CDM cosmology with $H_0$ = 70 km s$^{-1}$ Mpc$^{-1}$, $\Omega_{\rm m}$ = 0.3 and $\Omega_{\Lambda}$ = 0.7. All magnitudes are in the AB system.  

\begin{figure*}[!ht]
    \centering
\includegraphics[width=18cm]{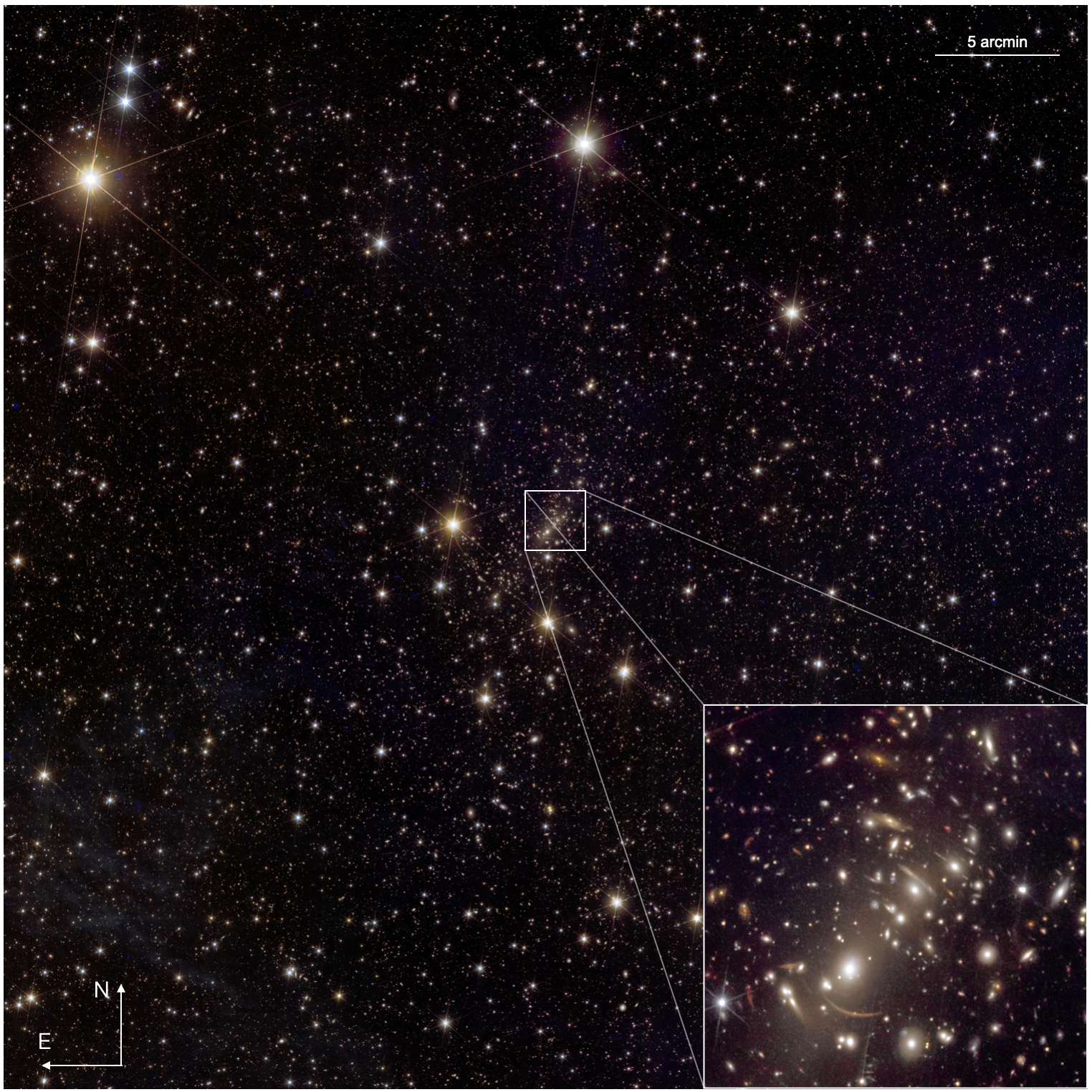}
    \caption{\euclid\, color composite image of the A2390 field. The blue channel is the $\IE$ image; green is \YE, and red is \HE. The whole field covers about 0.5\,deg$^{2}$. The inset is centered on A2390 and covers \ang{;2;}$\times$ \ang{;2;}; which is approximately the field of view of the Wide Field Camera 3 (WFC3) NIR channel onboard HST. The whole \Euclid\, field identifies approximately 250\,000 sources in the \texttt{CHI\_MEAN} detection image based on the \YE, \JE, and \HE coadd mosaic.}
    \label{fig:A2390_color}
\end{figure*}
\begin{figure*}
    \centering
    \includegraphics[width=18cm]{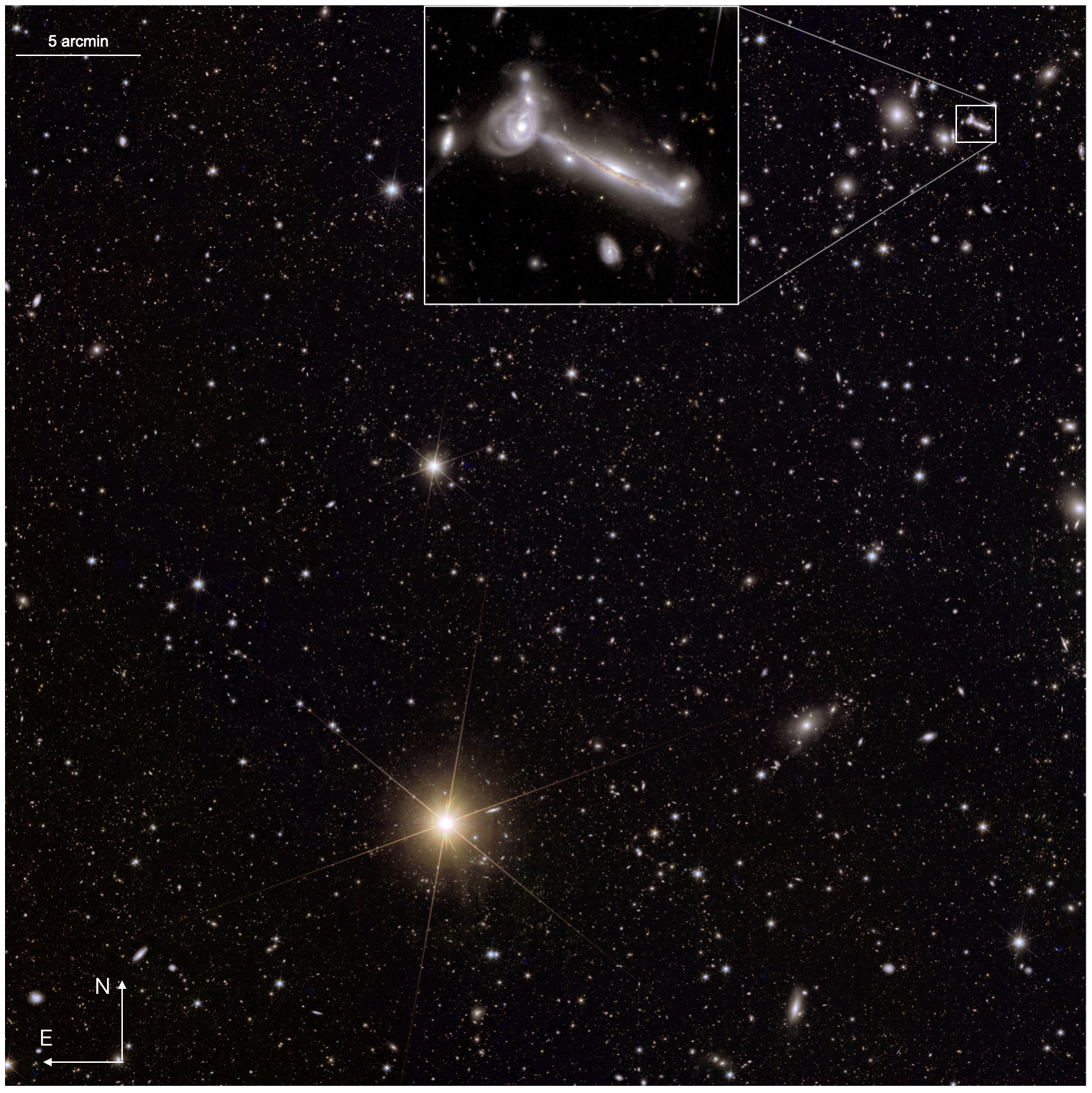}
    \caption{Color-composite image of the A2764 field using the $\IE$ image for the blue channel, \YE for green, and \HE for red. The image field of view is about 0.5 deg$^{2}$ and contains nearly 250\,000 objects based on the NISP coadd detection image. The inset shows a closeup of edge-on and face-on galaxies in the A2764 cluster.}
    \label{fig:A2764_color}
\end{figure*}

\section{The ERO program}
\label{sec:ero}

The ERO program was designed to obtain a comprehensive dataset spanning across multiple interdisciplinary working groups, including, but not limited to, those focused on the primordial universe, galaxy evolution and active galactic nuclei (AGN), strong and weak lensing, and galaxy cluster physics. These observations provide a first glimpse of the \euclid\ imaging data, with the aim of facilitating the inaugural evaluation of tools, data quality, and methodologies developed over the preceding decade. The early investigation of the dataset will help reveal and address challenges and limitations for a full scientific exploitation of the prime mission survey data. 

The ERO observations for this project were taken on 28 November 2023, as part of the performance-verification (PV) phase of the \Euclid\, mission.

\begin{figure*}[!ht]
    \centering
    \includegraphics[width=0.48\textwidth]{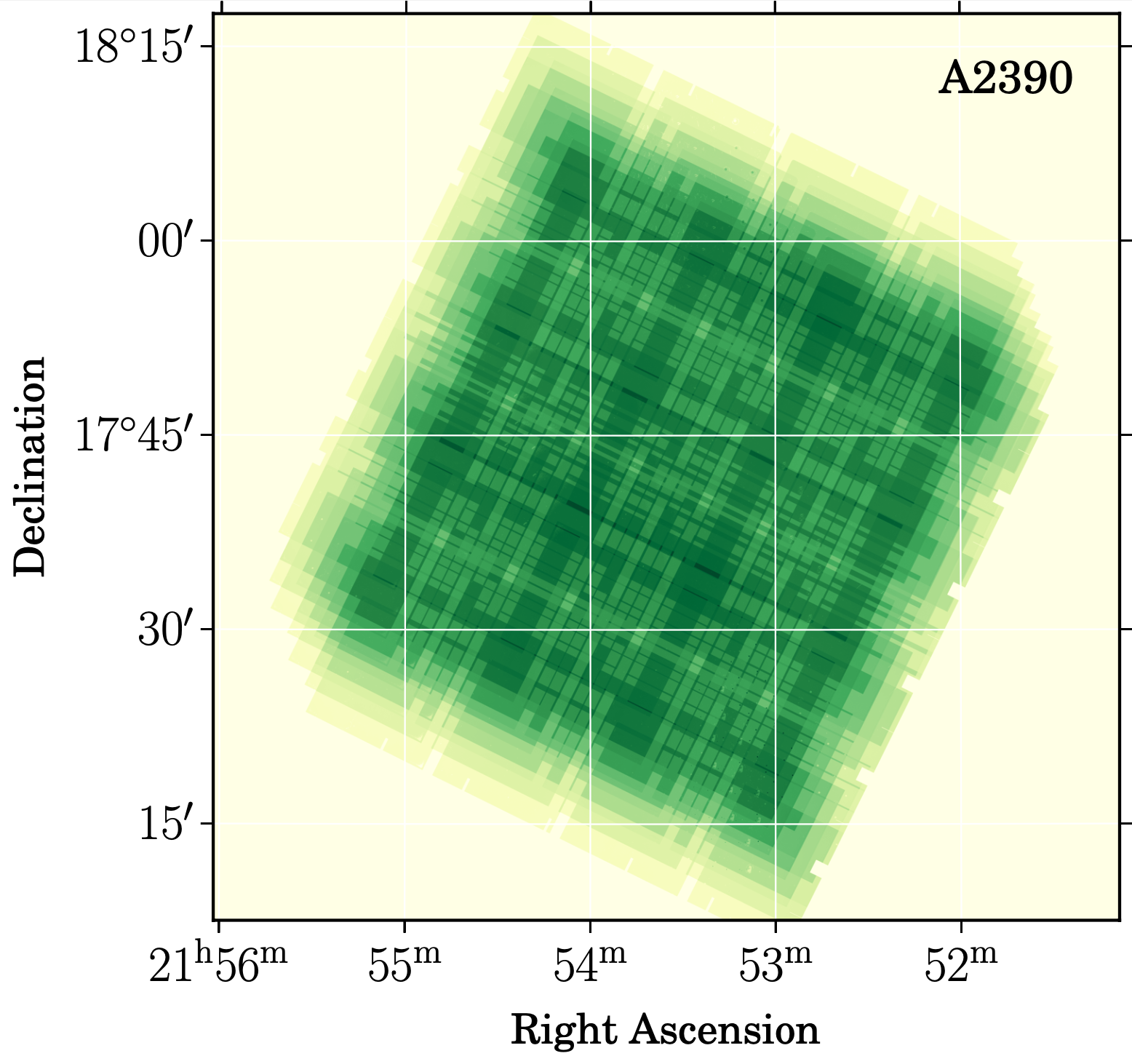}
    \includegraphics[width=0.5\textwidth]{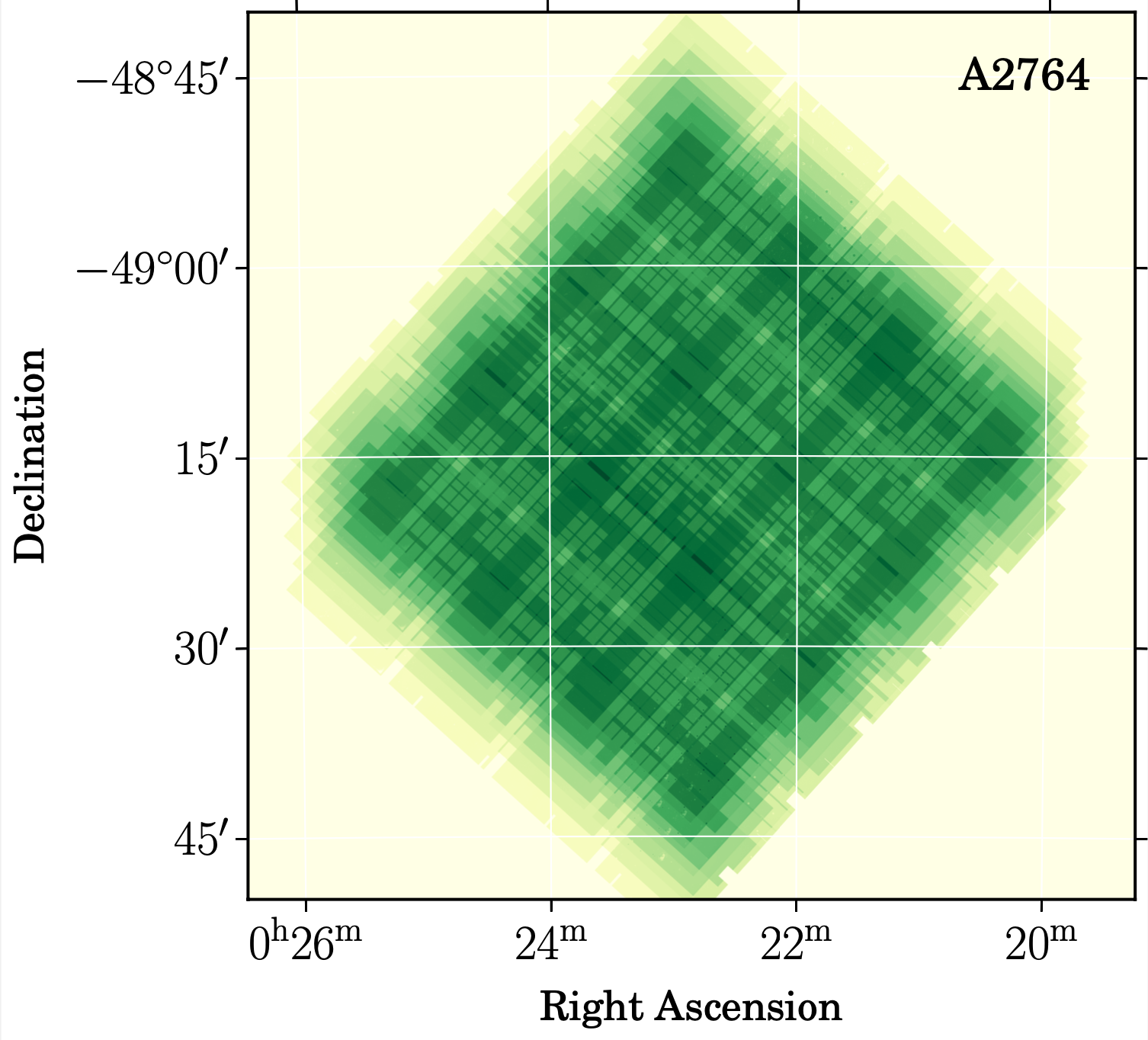}
    \caption{Weight maps of the \euclid\, observations of the lensing clusters A2390 (left) and A2764 (right). These examples are associated with the \JE\, observations. The maps demonstrate the large dither pattern adopted for these ERO observations and the sky coverage of each field.}
    \label{fig:weights}
\end{figure*} 

\subsection{Abell 2390} 

The first ERO target is the galaxy cluster A2390, RA=\ra{21;53;35.50}, Dec=\ang{17;41;41.54}, located at a redshift of $z=0.228$ \citep{sohn20}. It is one of the richest clusters in the Abell catalog \citep{pello91,leborgne91,feix10} with an estimated mass $M_{200}$\footnote{$M_{200}$ is the mass within the virial radius, which is where the mass density of the system is 200 times the critical density of the Universe.} ranging from $1.53 \times 10^{15} \, M_{\odot}$ from weak lensing \citep{okabe16} to $1.84 \times 10^{15} \, M_{\odot}$ from the projected phase-space of galaxies \citep{sohn20}. It includes multiple well-studied lensing arcs \citep[e.g.,][]{olmstead14,Richard21}. It also displays bright X-ray emission with a luminosity of  $2.7\times10^{45}\, {\rm erg\, s}^{-1}$ in the 0.1--2.4\,keV range \citep{bohringer98}. The \Euclid\ pointing is centered on the cluster, which represents a small fraction of the total area (cf. Fig. \ref{fig:A2390_color}).

\subsection{Abell 2764} 
The second ERO target is a random field on the sky that happens to cover the galaxy cluster A2764 in a corner of the field (see Fig. \ref{fig:A2764_color}). The coordinates of the center of the field are RA=\ra{00;22;50.1347}, Dec=\ang{-49;15;59.87}. The galaxy cluster was first identified by \citet{abell89}. It is located at a redshift of $z=0.07$ \citep{katgert96}, and is less massive ($M_{500}=9.2 \times 10^{13} \, M_{\odot}$) than A2390 \citep{sereno17}. It is also far less studied and characterized than A2390. Additional clusters can be identified in the field and are described in Sect.~\ref{sec:erosita}.

\section{Observations and data reduction}
\label{sec:obs}
\subsection{\euclid\ observations}
For each of the Abell clusters, we obtained three reference observing sequences \citep[ROS;][]{Scaramella-EP1}, implying 3 times the nominal \euclid\, Wide Survey exposure time. The duration of each ROS is about 70.2 min. A standard 4-point dither pattern, with an offset of 120\arcsec\, $\times$ 220\arcsec, was applied for each ROS. In addition, we adopted a 3\arcmin $\times$ 3\arcmin\ dither between the individual ROS to maximize the depth. The dithered pointings can be seen in the weight maps of the NISP imaging presented in Fig. \ref{fig:weights}. The large dithers fill the detector gaps and allow us to obtain continuous coverage of the field. The field of view of each combined mosaic is about 0.75\,deg$^{2}$.  

The data reduction process is described in detail in \citet{EROData}. Individual frames were stacked following two different approaches. We produced compact-sources stacks with the {\tt AstrOmatic SWarp} software \citep{bertin2002} using the individual multiple extension fits (MEF) files and the associated weight maps. The procedure achieves an astrometric accuracy of 5.1/15.2 milliarcseconds (mas) in \IE/\HE, respectively, for A2390 and 7.9/15.8 mas for A2764. A mesh size of 64 pixels (with a smoothing factor of 3) was adopted to subtract the background in input images that feature large extended low-surface brightness (LSB) emission. The second flavor of the data reduction also uses {\tt SWarp} on the individual frames and does not perform any internal background subtraction, thereby conserving LSB features such as interstellar cirrus emission. This is particularly useful for studies of diffuse intracluster light. The sigma-clipping algorithm used to build these extended-emission stacks slightly affects the photometry of very compact sources, such as stars, causing an additional 5\% photometric error in this case. The images were masked to contain only overlapping regions in the dithered frames of both VIS and NISP. The final sampling of VIS and NISP images is 0\arcsecf1 pix$^{-1}$ and 0\arcsecf3 pix$^{-1}$, respectively. A composite color image of A2390 is shown in Fig.~\ref{fig:A2390_color}. One can easily appreciate the gain in area, with one pointing only, over the limited survey capabilities of HST. The zoom-in inset illustrates the HST/WFC3 field of view over the center of the field (the lensing cluster). The total single pointing area of 0.75 deg$^{2}$ (cropped to 0.5 deg$^{2}$ in the color image) is larger than that of the entire Cosmic Assembly Near-infrared Deep Extragalactic Legacy Survey (CANDELS) area \citep{koekemoer11}, one of the largest HST extragalactic legacy programs. We also show a composite color image of A2764 in Fig. \ref{fig:A2764_color}.

\subsection{Ancillary observations}

These two ERO fields benefit from various space-based and ground-based ancillary data. For A2390, there is 6-band ($B,V,R_c,i,I_c,z'$) imaging data \citep{Miyazaki02} obtained with the Suprime-Cam instrument on the 8.2-m Subaru telescope. The data, which were reduced by the processing pipeline {\tt SDFRED} \citep{Yagi02,Ouchi04}, cover a 28\arcmin\, $\times$ 34\arcmin\, region centered on the brightest cluster galaxy \citep[BCG,][]{okabe16}. 
In addition, we also consider Canada France Hawaii Telescope (CFHT) WIRCam $J$-band observations, obtained in April-May-September 2010 covering the inner $24\arcmin\times 24\arcmin$ region, and a CFHT Megacam $u$-band (first generation MP9301) image taken in August 2007 for a total integration time of 2000\,s and covering the whole ERO field of view. Details of the reduction, astrometric and photometric calibration of these ancillary data will be presented in a subsequent paper dedicated to weak lensing and photometric redshifts. 

Previously, the cluster was imaged with HST using both the Wide Field Planetary Camera 2 (WFPC2) in F555W and F814W, and the Advanced Camera for Surveys (ACS) in F850LP \citep[see][]{Ota2012}. There are also VLT/MUSE integral-field spectroscopy (PID: 094.A-0115), which support the reported strong lensing mass modelling presented in \cite{Richard21}.   

In addition, the cluster core of A2390 has been observed with ALMA Bands 3-5 (PID 2021.1.00766.S) and SITELLE on the CFHT \citep[see][and references therein]{Qing2021}, which reveal an H$\alpha$ emission line in a cluster member galaxy. Furthermore, the cluster core of A2390 has been mapped with the SCUBA-2 submillimeter camera by \citet{Cowie22} to $1\sigma$ depths of $0.5$\,mJy at $850\,\mu$m and 2.5--3.5 mJy at $450\,\mu$m. Of the 59 submillimeter galaxies in the field, $44$ have a plausible cross-identification (within 2\arcsec) in the ERO data.

Both fields have been observed with the Dark Energy Camera \citep[DECam;][]{Flaugher15} on the Blanco telescope at the Cerro Tololo Inter-American Observatory. A2764 lies within the Dark Energy Survey \citep[DES;][]{Flaugher05, DES16} footprint, with typical $10\sigma$ depths ($1\arcsecf95$ aperture) for the $griz$-bands in DR2\footnote{\url{https://des.ncsa.illinois.edu/releases/dr2}} of 24.7, 24.4, 23.8 and 23.1, respectively \citep{DESetal}. Typical image qualities for these DES bands are $1\arcsecf11$, $0\arcsecf95$, $0\arcsecf88$ and $0\arcsecf83$. DES data will be used to partner the Southern regions of the Euclid Wide Survey during {\em Euclid}'s first data release, and so this largely-blank field provides an early example of those forthcoming combined data. Meanwhile, A2390 has been observed in $grz$-bands as part of the DESI Legacy Survey \citep[DLS;\footnote{\url{http://legacysurvey.org/}}][]{DLS2019}. Typical depths ($5\sigma$, extended source) and image qualities of the DLS images are, 23.72, 23.27, 22.22 and $1\arcsecf29$, $1\arcsecf18$, $1\arcsecf11$, in the $grz$-bands, respectively.

\begin{table}
    \centering
    \small
      \caption{Limiting magnitudes at $5\,\sigma$ reached in each \Euclid\ filter of the A2390 and A2764 fields. The depth measurements assume an aperture diameter of 0\arcsecf6 and 0\arcsecf3 for NISP and VIS, respectively.}
    \begin{tabular}{l c c c c c c}
       Field  &RA   & Dec    &  \IE & \YE & \JE & \HE \\
              &[deg]&[deg]   &      &     &     &      \\ 
       \hline\\
       A2390  & 328.397 &$+17.709$  & 27.01 & 25.18 & 25.22 &25.12 \\
       A2764  & 5.713   &$-49.249$ & 27.26 & 25.30 & 25.41 &25.21  \\
       \hline\\
    \end{tabular}
    \label{tab:depth}
\end{table}


\section{Photometric catalogs}
\label{sec:catalogs}

\begin{figure*}
    \centering
    \includegraphics[width=8.5cm]{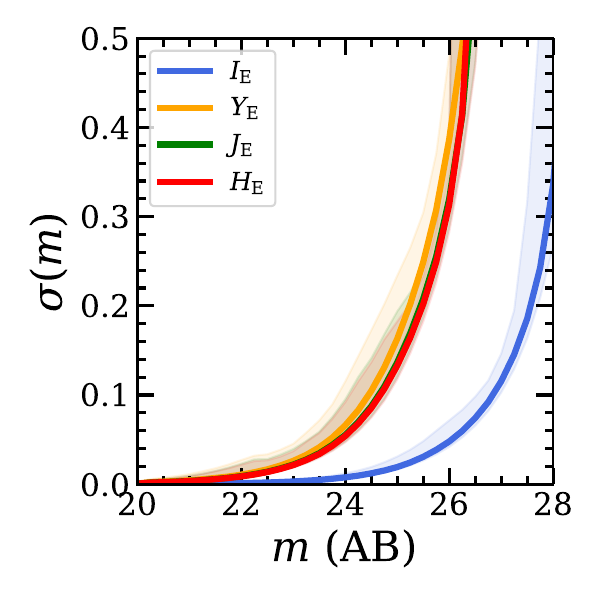}
    \includegraphics[width=8.5cm]{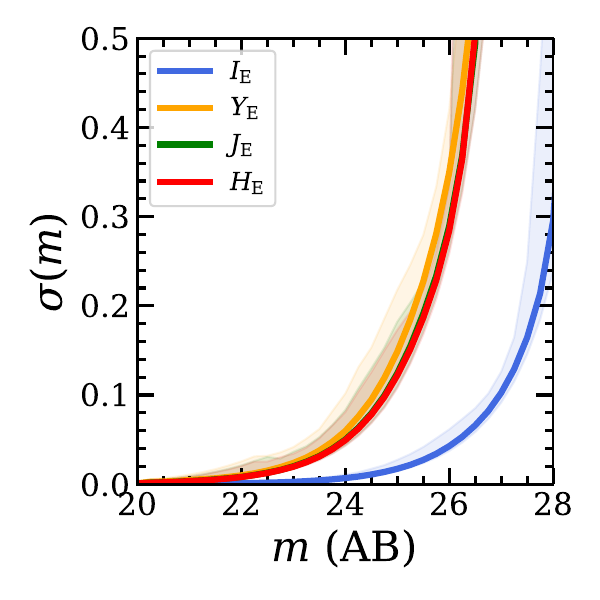}
    \caption{Photometric errors as a function of the magnitude for all sources in the catalog for A2390 (left) and A2764 (right). Each curve represents the median error for each VIS and NISP band, while the corresponding shaded area is the 1-$\sigma$ dispersion. The error measurements are described in Sect. \ref{sec:sep}.}
    \label{fig:magerr}
\end{figure*} 

\subsection{Point-source aperture photometry} 
\label{sec:sep}
The combination of NISP NIR bands (\YE, \JE, and \HE) and ultra-deep VIS optical imaging in $\IE$, about 2 magnitudes deeper, enables immediate searches for ultra-red objects with no appreciable optical flux. Such objects include high-redshift galaxies and quasars ($z\gtrsim6$) as well as dusty and/or quiescent galaxies, cool stars and brown dwarfs at lower redshifts. The former were a key scientific goal of this ERO program. To enable rapid searches for such objects \citep[see][]{EROLensVISdropouts}, we first co-add the compact-source version of \YE+\JE+\HE\, mosaics to produce a \texttt{CHI\_MEAN} detection image using \texttt{SWarp} \citep{bertin2002}. From that image, we detect 501\,994 objects (249\,319 in A2764 and 252\,675 in A2390) using Pythonic SourceExtractor \citep{bertin96, barbary16} assuming a threshold of $1.5\,\sigma$ within at least three contiguous pixels of the detection image smoothed by a Gaussian kernel with a 0\arcsecf45 (1.5 pixels) FWHM.\footnote{A discussion on thresholds and kernel sizes will be presented in Zalesky et al. in prep.}. De-blending is configured with $2\times10^{5}$ thresholds requiring a contrast of $10^{-5}$. No cleaning is applied as it is liable to remove real sources of interest. Photometry is extracted at the source positions by summation of pixels within several circular apertures with diameters: 0\farcs1, 0\farcs2, 0\farcs3, 0\farcs6, 1\farcs0, 1\farcs2, 1\farcs5, and 2\arcsecf0. Uncertainties are estimated strictly from extractions on the weight map, whose distribution in effective per-pixel error matches well to the observed noise level in the images. Given the relative similarity of NISP PSFs measured by stacking point sources in the images, we elect to simply correct all apertures to total flux, assuming point-like light distributions outside the apertures. Although we provide similarly corrected total photometry for VIS, it is the signal-to-noise ratio that is most important to our objective, where aperture corrections are superfluous. Therefore, given the uncertainty in recovering total fluxes, in particular for extended objects, we strongly caution against combining this photometry with other pre-compiled catalogs as colors are not guaranteed to be robust. Figure \ref{fig:magerr} shows the typical magnitude uncertainty in each band as a function of magnitude for all the sources detected and included in the photometric catalog. It highlights, in particular, the stable photometry across a wide flux range and the significant difference in depth between the VIS and the NISP observations. Magnitude zero point uncertainties are estimated to be below a 10\% level in these data. We estimated the limiting magnitude for each of the \euclid\ bands of our observations by measuring the flux in 10\,000 empty apertures randomly distributed in the A2390 and A2764 images. The standard deviation of the flux distribution provides the typical depths quoted in Tab. \ref{tab:depth}. 

We provide a \IE,\YE,\JE,and \HE\, photometric catalog of the two ERO fields, which includes unique and field-specific IDs, coordinates, basic shape parameters, aperture and total fluxes, aperture-to-total corrections, and {\tt SourceExtractor}-like flags. All photometric measurements are corrected for Galactic extinction  using the extinction curves of \citet{Fitzpatrick2007}. A total of 73\,841 unreliable sources (14.7\% of the total) likely to be artifacts near mosaic edges (where only a single dither contributes to the mosaic) or near bright stars are flagged by \texttt{use\_phot}=0. See the accompanying \texttt{README} file for details of the source detection, photometry, and catalog contents. A case study utilizing this catalog to identify NISP-only sources is presented in \citet{EROLensVISdropouts}.

\subsection{SourceXtractor++}
\label{se:SE++}
We separately derive photometric catalogs containing morphological parameters based on the recent re-implementation of {\tt SExtractor} \citep{bertin96}, called {\tt SourceXtractor++}  \citep[hereafter {\tt SE++};][]{Kuemmel22,Bertin22}. Among several other features, this code allows the user to fit PSF-convolved parametric models describing the surface brightness profile of every single source in the images, the fitting being performed simultaneously on several exposures grouped into a number of photometric filters or bands. The PSF, and its spatial variations across the field of view, are described with the {\tt PSFEx} tool \citep{Bertin11}. In the framework of the Euclid Morphology Challenges \citep{Merlin-EP25,Bretonniere-EP26}, we have shown that {\tt SE++} can recover accurate multi-band photometry and morphological parameters, such as half-light radius, ellipticity, orientation, and S\'ersic index. \citet{Great3-Mandelbaum15} and \citet{EP-Martinet19} have previously shown that recovered shapes obtained with the model-fitting engine of {\tt SExtractor} meet the weak lensing standards (with a proper weighting of source ellipticities) that is sufficient for the scopes of cluster analysis and stage III weak lensing surveys. A forthcoming weak lensing study of A2390 will investigate in detail the merits of weak lensing shape derivations with three methods: SE++, \textsc{LensMC}, currently the main cosmic shear measurement adopted for the first data release in the OU-SHE part of the \euclid\, Ground Segment (see Congedo et al. 2024, in prep.), and {\tt KSB++} \citep{kaiser1995,luppino1997,hoekstra1998}.

We ran {\tt SE++} in two settings, but source detection was always performed in the VIS \IE~band. In the first run, only the VIS band was used to constrain a single S\'ersic profile. This provides shapes that can readily be used for weak lensing, and best-fit sizes provide a complementary star/galaxy discriminator.
Then, a two-component bulge+disk (i.e. de Vaucouleurs + exponential) concentric model was also fit to all the ground-based, VIS and NISP bands at once, i.e. $u, B, V, Rc, i, Ic, z, \IE, \YE, \JE, \HE$. The orientation of the  disk and the bulge was forced to coincide, but their axis ratios were left free. The bulge and disk half-light radii were constrained to be constant across wavelengths, but the bulge-to-total flux ratio was allowed to vary. Furthermore, in order to account for the long time span between SuprimeCam (early 2000') and {\Euclid} (November 2023) observations, we allow for proper motions for all sources.

Both runs rely on the same PSF determination based on {\tt PSFEx}. In addition, further measurements were performed besides those that depend on model fitting. In particular, aperture photometry in 1$\arcsec$, 3$\arcsec$, and 6$\arcsec$ diameter apertures was computed, as well as source centroid and isophotal moments, in line with standard {\tt SExtractor} measurements.

\begin{figure}[!h]
    \centering
    \includegraphics[width=0.49\textwidth]{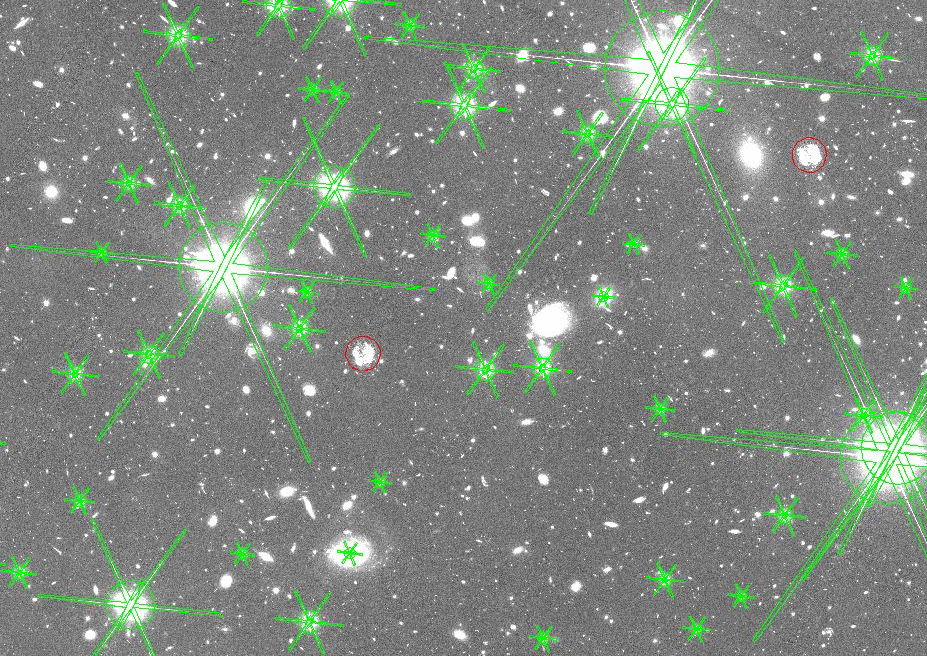}
    \caption{\euclid\, \IE~image ($1\farcm5\times1\farcm0$) in the A2390 field of view, showing masked area (inside polygons). the image shows diffraction spikes, saturation bleeds, core region, and two reflection ghosts of 8$\arcsec$ radius (in red).}
    \label{fig:vis_masks}
\end{figure}

We applied a careful masking of sources in the vicinity of bright stars with an automated scaling with stellar magnitude of the size of polygons, aimed to mask out the extended diffraction spikes (Fig. \ref{fig:vis_masks}). A model of ghosts due to reflections onto the dichroic, which splits light between the VIS and NISP instruments, was also defined. Ghosts from Gaia-DR3 stars brighter than $G_{\rm RP}=12$ are automatically flagged as a disk of $8\arcsec$\, radius, shifted by about $1\arcmin$ with respect to the source star. Likewise, stars (core and spikes) between Gaia magnitudes $G_{\rm RP}=8.5$ and $G_{\rm RP}=18.5$ are automatically masked. Since the brightness profile along diffraction spikes drops with the angular distance to the star center $\theta$ as $\theta^{-2.2}$, we scale the size of corresponding masks as $\exp(-0.41~ G_{\rm RP})$. After automated generation of masks, further tuning was done by hand.  In the case of A2390, before masking, the total relevant area is 0.75 deg$^2$. After masking out those regions, the useful VIS area reduces to 0.71 deg$^2$. The unmasked area with deep multi-band SuprimeCam data further reduces to 0.21 deg$^2$.

\begin{figure*}[!h]
    \centering
    \includegraphics[width=0.49\textwidth]{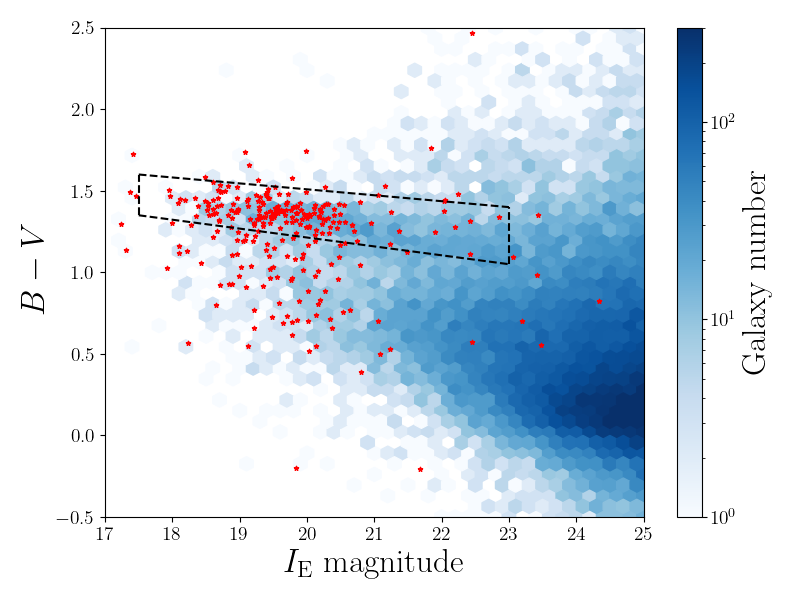}
    \includegraphics[width=0.49\textwidth]{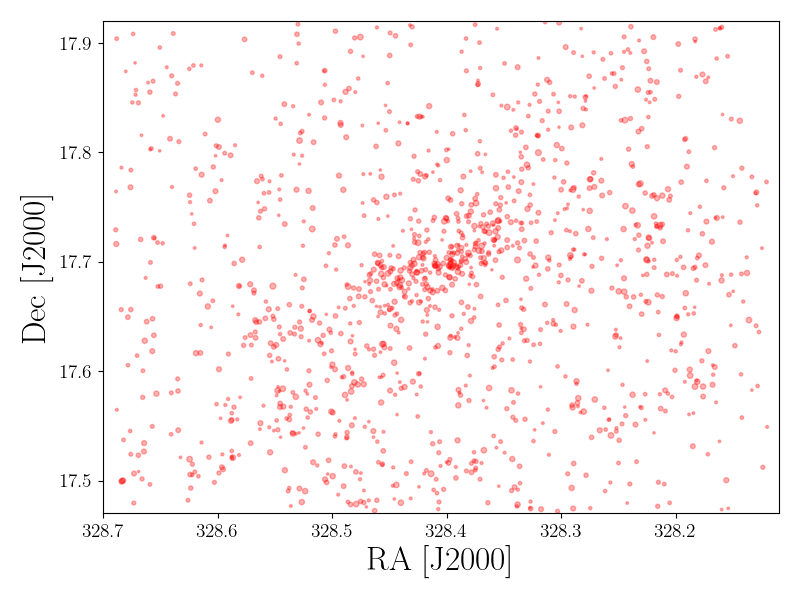}
    \caption{{\em Left:} Galaxy number density in the plane of Subaru SuprimeCam $B-V$ color vs. {\em Euclid} \IE-band magnitude, from bulge+disk light profile fits with \texttt{SourceXtractor++} (see text). Red stars indicate spectroscopic cluster members of A2390 determined by \cite{sohn20} and MUSE observations of the core of A2390. The dashed region indicates the cluster red sequence based on the spectroscopic members and extrapolated to fainter magnitudes. {\em Right:} Sky location of galaxies enclosed by the red-sequence selection in the left-hand panel over the joint $0.21~$deg$^2$ area covered by \Euclid\ and Subaru SuprimeCam. The size of each point scales inversely with $\IE$ magnitude to indicate the distribution of stellar mass in red galaxies at the cluster redshift.}
    \label{fig:red_sequence}
\end{figure*}

Multi-band photometry at fixed positions allows us to reach greater depth in the filters that are not the VIS \IE~detection image. We reach a 0.1 magnitude uncertainty on model magnitudes for A2390 (total bulge+disk decomposition) at 24.14, 26.19, 25.80, 26.41, 25.48, 25.64, and 24.76 for filters $u$, $B$, $V$, $R_c$, $i$, $I_c$, and $z$, respectively. This depth, well-matched to \Euclid\ depth, was achieved thanks to exquisite observing conditions, during which seeing reached $0\farcs905$, $0\farcs736$, $0\farcs572$, $0\farcs561$, $0\farcs516$, $0\farcs682$, and $0\farcs570$, respectively. Once combined with VIS and NISP \Euclid\ photometry, this will enable computation of
accurate photometric redshifts, as will be shown in the dedicated weak lensing paper. As an illustration, note the tight red sequence of cluster member galaxies in the left panel of Fig.~\ref{fig:red_sequence}. For details, refer to the \texttt{README} file associated with the catalog release. 

\begin{figure*}[!h]
\begin{center}
\includegraphics[width=14cm]{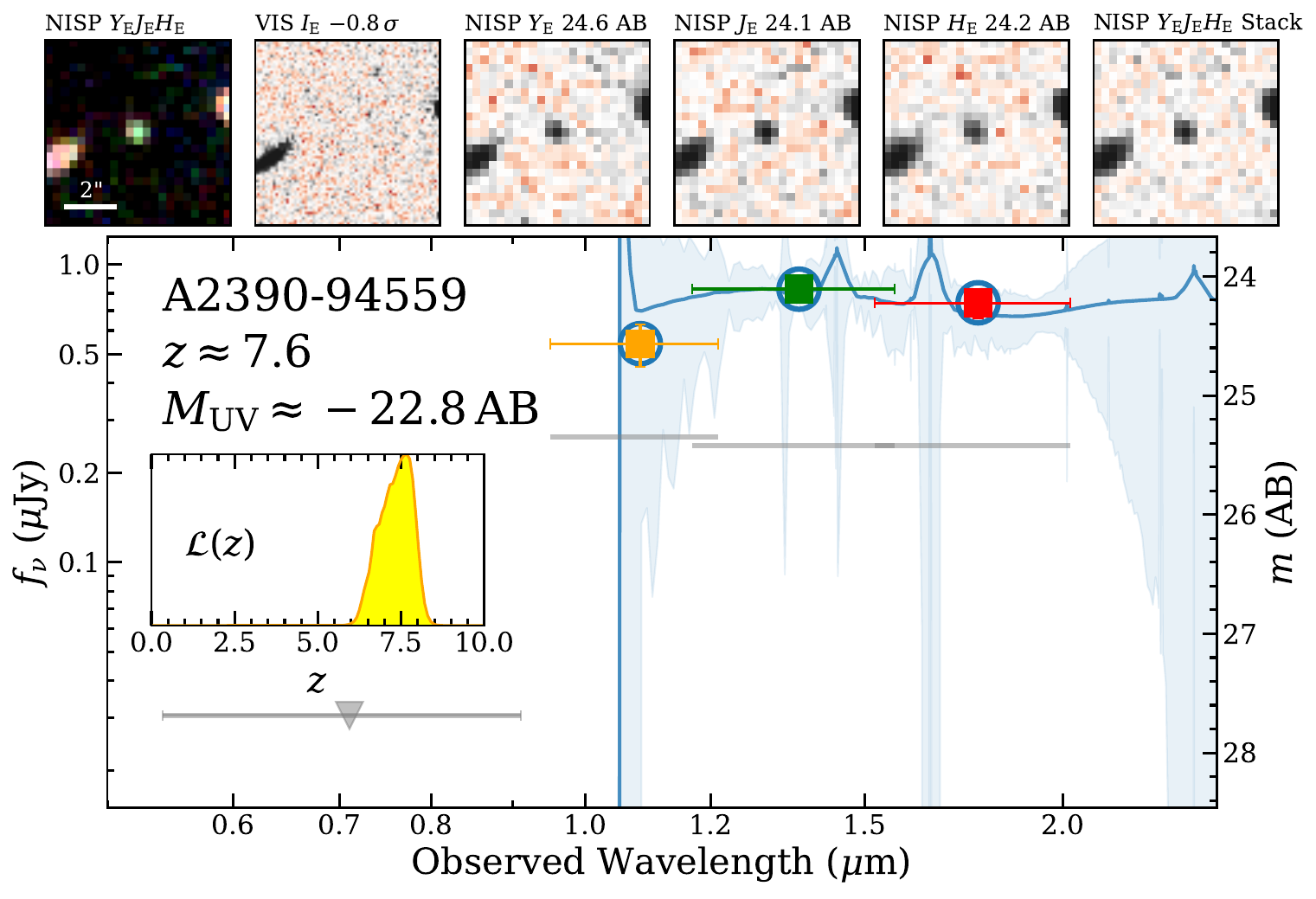}\\
\includegraphics[width=14cm]{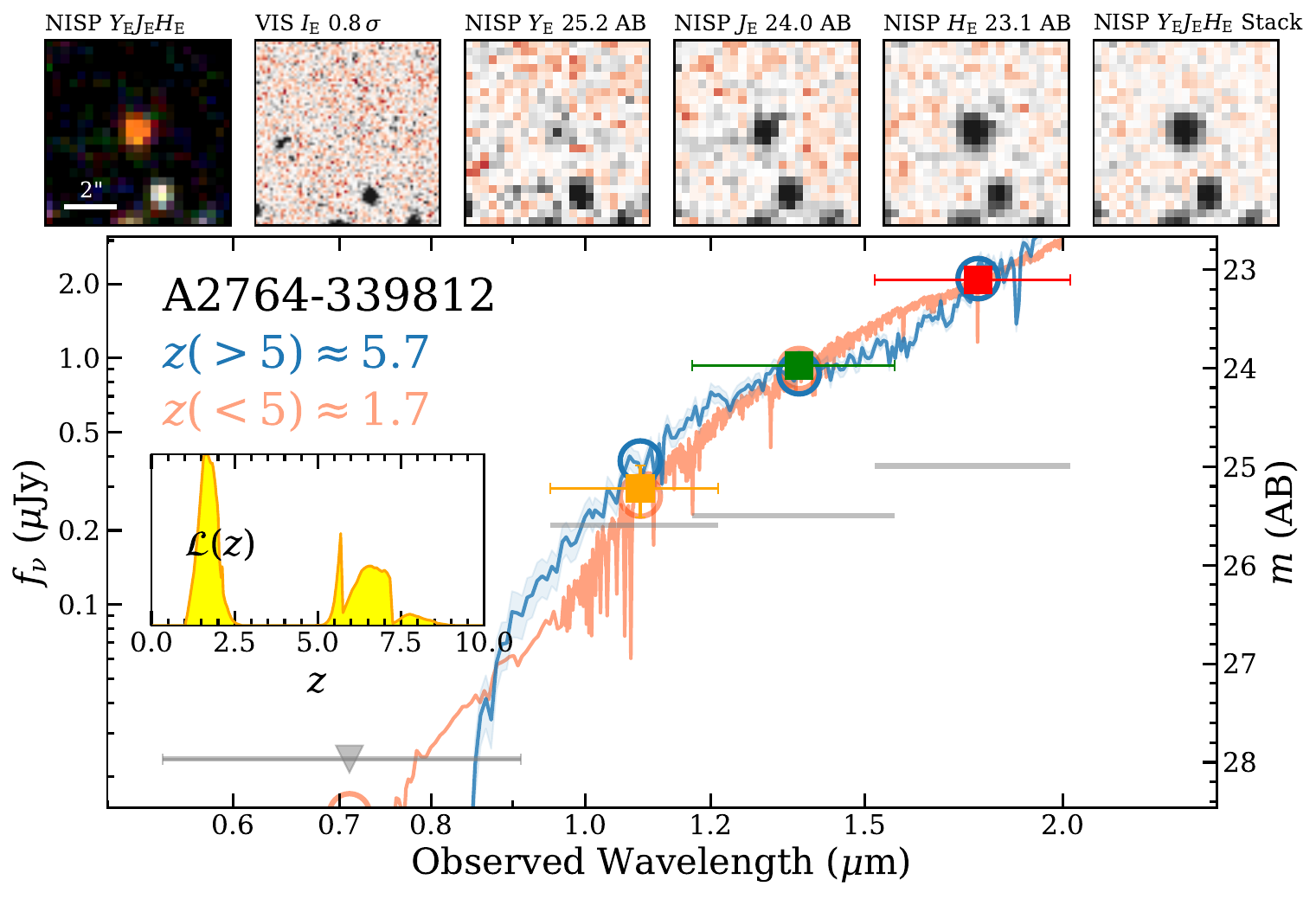}
\end{center}
\caption{The identification of VIS-dropout candidates in the A2390 and A2764 fields. The {\bf top panel} shows an example of an LBG candidate at a redshift of $z>6$, while the {\bf bottom panel} shows an ERS, most likely located at a lower redshift. For each of these sources, the top row shows the cutout images of the source in each VIS and NISP filters, together with a NISP stack and a color-composite image based on the \YE, \JE, and \HE\ filters. The size of these images is about 7\arcsec\ $\times$ 7\arcsec. The bottom figure illustrates the best-fit model (blue and orange curves) to the \euclid\ photometry (colored circles with associated $1\,\sigma$ uncertainties). The gray horizontal bars denote the $3\, \sigma$ depth of each band, while the down-pointing triangle marks the upper limit of the \IE\ band. The left inset presents the posterior distribution function for the photometric redshift \citep[see][for details]{EROLensVISdropouts}.}
\label{fig:dropouts}
\end{figure*}

\section{High-redshift dropout sources}
\label{sec:dropouts}
The relative depths of VIS and NISP imaging, which together cover the entire 0.55--2 $\mu m$ spectral range, means we can identify high-redshift ($z>6$) dropout sources, that are well detected in NISP and remain undetected in the deeper VIS data. Here we present a summary of the results from the selection of high-redshift sources. A detailed analysis of these findings, along with the challenges in identifying high-redshift candidates, is provided in a dedicated paper \citep{EROLensVISdropouts}. We achieve a $5\,\sigma$\, depth in the range 25.1--25.4 in the NISP bands, and 27.1--27.3 in the VIS image, assuming circular apertures with diameter 0\arcsecf6 and 0\arcsecf3, respectively (see Table \ref{tab:depth}). In addition, we require our sources to be detected at a $3\,\sigma$\, level in the $\YE$ band and to be simultaneously undetected in the $\IE$ band at a $1.5 \sigma$ level. This leads to a source selection that is based on a minimum Lyman break of approximately 2 magnitudes, which is significantly higher than most Lyman-break selection criteria used in HST, and JWST, identification of $z>6$ galaxies \citep[e.g.,][]{ bouwens15,finkelstein15,Atek_2023b}. A stronger dropout criterion significantly reduces contamination by red low-redshift interlopers. However, ERSs can still enter our selection window but can be differentiated from LBGs by their clearly extreme red colors compared to the relatively flat or blue UV slopes observed in LBGs above the Lyman break (cf. Fig. \ref{fig:dropouts}) and the existence of a secondary peak at low redshift in the posterior probability distribution of their photometric redshift (see below). The ERSs are an interesting population that constitutes its own science case at lower redshift. The strong Lyman break does not completely exclude some cool brown dwarf stars, especially given the limited spatial resolution of NISP, which only enables resolving multi-component bright galaxies, which might be common at high redshift \citep{bowler17}. To minimize contamination by artifacts, we also require the sources to be compact by imposing a flux ratio between two apertures of \JE(1\arcsecf5)/\JE(0\arcsecf6) < 1.3. We have verified that the criterion does not exclude real sources from the selection. Furthermore, in order to mitigate contamination by persistence at NIR wavelengths from previous exposures, we performed a time series analysis on the individual frames. We removed all sources that showed a typical declining light curve \citep[for more details, see][]{EROLensVISdropouts}. 

This color selection is complemented with photometric redshift estimates using the spectral energy distribution (SED) fitting tool {\tt eazy-py} \citep{Brammer2008, brammer22}. We assume a redshift range of $0.01<z<15$ and a flat luminosity prior using the {\tt agn\_blue\_sfhz} set of galaxy templates. All the sources have a posterior distribution function (PDF) that is consistent with the $z>6$ solution, with no significant secondary solution at low redshift. The full procedure and the dropout sample are described in detail in \citet{EROLensVISdropouts}. Combining the two ERO fields, we identify a total of 30 LBGs and 139 ERSs, from a total of about 500\,000 sources. The absolute magnitudes of the LBGs ranges between $M_{\rm UV}=-21.9$\, and $M_{\rm UV}=-23.6$. In Fig. \ref{fig:dropouts}, we show an example for each of these categories. The top row shows the photometry in each of the NISP and VIS filters and a composite color image of the \YE, \JE, \HE\, images. We also show the best-fit solutions of the photometric data derived with {\tt eazy-py}, and the associated probability distribution. The LBG candidate has a clear high-redshift solution centered at $z\sim7.7$, whereas, the ERS has a significant low-redshift solution around $z\sim1.7$.

These early results, based on a small portion of the sky relative to the full survey, highlight the crucial role that \Euclid\ will play in identifying the brightest galaxies during the epoch of reionization. Given the uncertainties on the exact shape of the bright-end of the UV luminosity function at $z\sim8$--10 \citep{bowler20}, on the one hand, and the apparent excess of UV-bright galaxies at $z>10$ \citep{harikane23,chemerynska24,adams23,donnan24}, on the other hand, \Euclid\ imaging and subsequent spectroscopic follow-up will be instrumental in constraining the exact abundance of bright high-redshift galaxies, and the exact shape of the bright end of the UV luminosity function at $z=$6--9. The cross-correlation with Ly$\alpha$ emitters will allow the identification of giant ionized bubbles, of the order of 10 Mpc in size, and map the ionization state of the IGM. In addition to bright LBGs, the \euclid\ data will also allow the identification of bright quasars in the deep and wide surveys. The ability of current data to distinguish between high-redshift quasars and galaxies remains limited in the absence of follow-up spectroscopy. \euclid\ grism spectroscopy will be limited to sources brighter than 22 mag. These limitations also apply to potential contaminating sources at low redshifts, such as cool brown dwarf stars \citep{EROLensVISdropouts}. Again, the relative depth of VIS imaging compared to NISP, which leads to very strong breaks between these two wavelengths, reduces the contamination rate of Galactic dwarfs, but will remain an important source of contamination \citep{Barnett-EP5}. Fortunately, the brightness of these sources makes them ideal candidates for follow-up spectroscopy from the ground in order to confirm their redshifts. Determining the number, density, and luminosity of these sources will motivate follow-up campaigns, with JWST for example, to constrain the nature of the brightest sources in the early Universe, including galaxies and quasars.

\section{Weak lensing analysis}
\label{sec:wl}
\begin{figure*}[!h]
\begin{center}
\includegraphics[width=9cm]{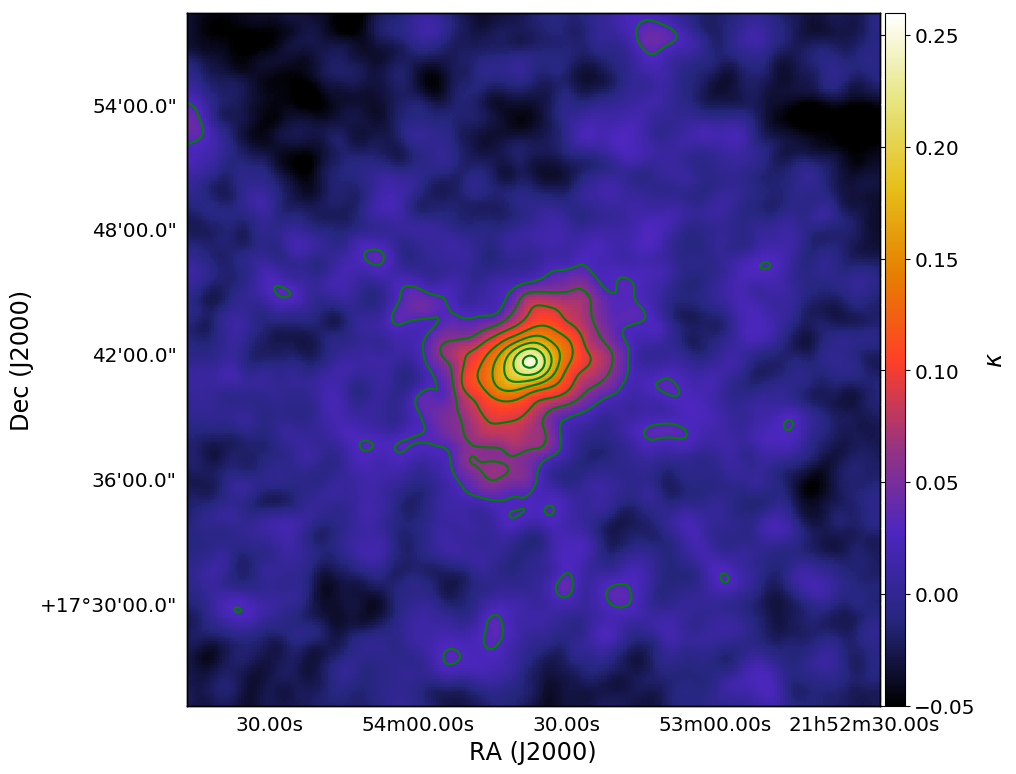}
    \includegraphics[width=9cm]{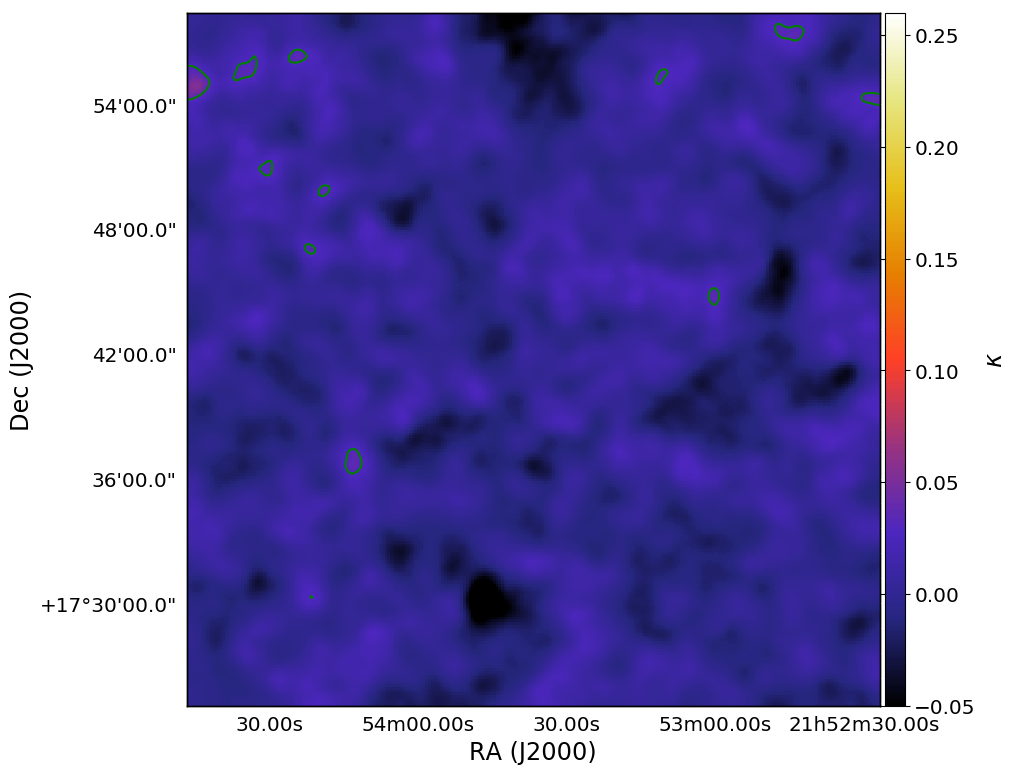}
\end{center}
\caption{WL convergence $\kappa$ reconstruction of the A2390 field. The left panel shows the reconstructed E-mode signal, with the cluster detected in the central region with high significance. The right panel shows the B mode, which does not include a cosmological signal to first order, but provides an illustration for the noise level of the reconstruction. The green contours start at $\kappa=0.03$ and are in steps of $\Delta\kappa=0.03$. Both reconstructions assume a zero mean convergence when averaged over the full reconstructed sky area.}
\label{fig:WLmap}
\end{figure*}

Weak gravitational lensing (WL) is one of {\euclid}'s primary cosmological probes. By measuring the shapes of background galaxies, WL analyses allow for the reconstruction of the projected foreground mass distribution \citep[e.g.][]{bartelmann01}, which is dominated by invisible dark matter. The WL signature of one of the targets of this ERO program, A2390,  has previously been studied using ground-based WL data \citep[e.g.][]{vonderlinden14,applegate14,okabe16}. However, given their outstanding depth and resolution, the new \euclid\, observations of the cluster provide an excellent opportunity to showcase {\euclid}'s WL capabilities. For this, we have measured WL galaxy shapes in the  A2390 VIS observations using different shape measurement algorithms including {\tt \textsc{LensMC}} (which is designated as the primary shape measurement of the {\euclid} Data Release 1), {\tt SE++} (see Sect.\thinspace\ref{se:SE++}), and {\tt KSB+} \citep{kaiser1995,luppino1997,hoekstra1998}. The details of this analysis will be presented in a separate paper. Here we briefly highlight first results from the {\tt KSB+} analysis, which employs the \citet{erben2001} implementation of the {\tt KSB+} algorithm, including the shear bias calibration from  
\citet{hernandez20}, as used, e.g., for the analysis of HST WL observations of distant galaxy clusters from the South Pole Telescope in \citet{schrabback18,schrabback21b} and \citet{zohren22}. After applying {\tt KSB+} shape measurement selections, removing masked objects in the vicinity of bright stars (see Sect. \ref{se:SE++}) and very extended galaxies, and requiring a flux signal-to-noise ratio ${\rm S/N}>10$ (defined as the \texttt{FLUX\_AUTO/FLUXERR\_AUTO} ratio from \texttt{SExtractor}), as well as a magnitude in the range $22<\IE<26.5$, the resulting catalog provides a WL density of 56.5 sources per arcmin$^2$ when averaged within the central 33\farcm3 $\times$ 33\farcm3 region around the cluster center. As an initial result of this analysis, we present a reconstruction of the convergence field $\kappa$  in Fig.\thinspace\ref{fig:WLmap}. The convergence maps the foreground mass distribution, but is scaled according to the geometric lensing efficiency \citep[e.g.][]{bartelmann01}. In particular, we apply a Wiener-filtered reconstruction following the procedures described in \citet{mcinnes09}, \citet{simon09}, and \citet{schrabback18}. Here, the signature of the cluster is detected with high significance in the curl-free E-mode reconstruction (left panel of Fig.\thinspace\ref{fig:WLmap}). In the right panel of Fig. \ref{fig:WLmap}, we additionally show a B-mode reconstruction derived from the curl component of the WL shear field. To first order, gravitational lensing only creates curl-free E modes. Accordingly, the B-mode reconstruction provides an assessment for the level of noise (and potential residual systematics) in the reconstruction. 

\section{Weak plus strong lensing analysis}
\label{sec:sl}
\begin{figure}
    \centering
    \includegraphics[width=0.48\textwidth]{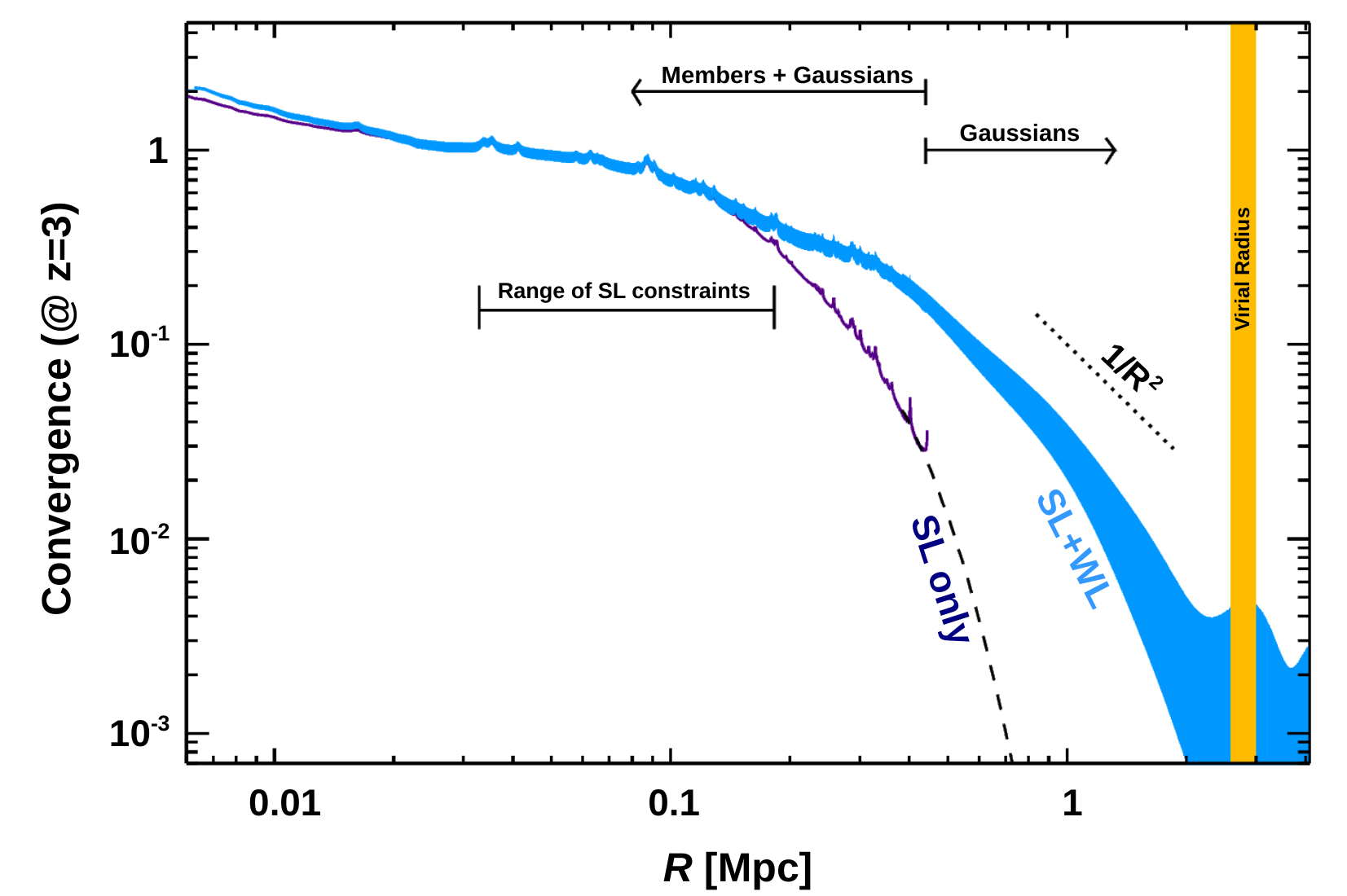}       
    \caption{Mass profile of A2390 from an SL-only analysis and a joint SL+WL analysis. The two regions that include members+Gaussians ($R\lesssim 0.44$ Mpc) in the lens model or only Gaussian functions are indicated. We also show the range of distances covered by the SL constraints (from the brightest cluster galaxy).  
    For the SL+WL combination, we show a range of models that are consistent with the data ($\approx 1$ standard deviation). 
    The addition of WL data from {\Euclid} is key to properly constrain the mass of the cluster out to the virial radius (corresponding to $R_{200}$ or radius at which the density reaches 200 times the mean density of the Universe) with a vertical orange bar, where the width is the uncertainty in the virial radius.}
    \label{fig:A2390_SLWL}
\end{figure}

\euclid\ WL data can be combined with existing strong lensing (SL) constraints (or new multiple-image systems identified by \Euclid) enabling a joint SL+WL analysis. SL constraints are useful at constraining the inner region of clusters (typically the central 300 kpc), but do not constrain the outer regions of clusters, thus limiting our ability to estimate their virial masses. {\Euclid}'s WL measurements can extend the analysis of the lensing signal to beyond the virial radius. When several multiply-imaged galaxies exist at different redshifts, the combination of both data sets breaks the mass-sheet degeneracy \citep[e.g.][]{schneider08}. Among the ERO targets, A2390 offers a unique opportunity to test the capacity of \Euclid\ at measuring the virial mass of a massive cluster. This cluster contains 11 multiply-imaged galaxies with confirmed spectroscopic redshifts \citep{Richard21}. These galaxies produce 23 multiple images that result in 46 SL positional constraints. We combine these constraints with WL measurements from \euclid. We rebin the WL data in $1\times1$ arcmin$^2$ bins.
The analysis is performed with the free-form code {\tt WSLAP+} \citep{diego07,sendra14}. This algorithm assumes the mass distribution can be decomposed as a superposition of 2D Gaussian  functions. The amplitude of each Gaussian is optimized by minimizing the difference between the observed and predicted lensing variables (mean shear in the adopted $1\times1$ arcmin$^2$ bins for the WL portion of the data set, and positions of multiply imaged galaxies for the SL portion of the data set). The unknown position of the multiply imaged galaxies in the source plane is also optimized together with the distribution of mass.  In the inner portion of the region constrained by the SL measurement, the light distribution of the most prominent member galaxies (approximately 70 for this cluster) is assumed to trace the mass on small scales with a given mass-to-light ratio, that is also derived as part of the optimization. The resulting convergence profile is shown in Fig.~\ref{fig:A2390_SLWL}, where we show the solution obtained when only SL data are used and when both SL and WL data are used. The profile in the inner region ($R\lesssim0.44$ Mpc) contains the  contribution from member galaxies. Even though member galaxies make only a small contribution to the profile of the convergence (except in the central peak), these galaxies are relevant for accurately predicting the positions of the multiply-imaged lensed galaxies and they also act as a regularization mechanism for the solution to converge \citep[see][]{diego07,sendra14}.   
The SL+WL profile differs significantly from the SL profile at radii beyond the region constrained by SL. This is a consequence of the SL-only solution not being constrained in this regime, where {\tt WSLAP+} solutions are known to be biased low \citep[this is a memory effect of the initial guess as described in][]{diego07}. In contrast, the SL+WL combination recovers a solution consistent with the expected Navarro-Frenk-White (NFW) profile \citep{navarro96} of massive clusters at large radii.

\section{\euclid's preview of high-redshift galaxy clusters detected in the field of A2764}
\label{sec:erosita}

\begin{figure}[htbp!]
\centering
\includegraphics[angle=0,width=1.01\hsize]{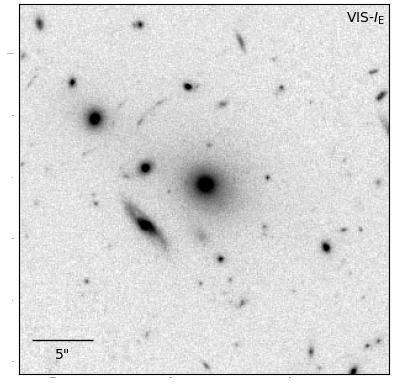}
\includegraphics[angle=0,width=1.01\hsize]{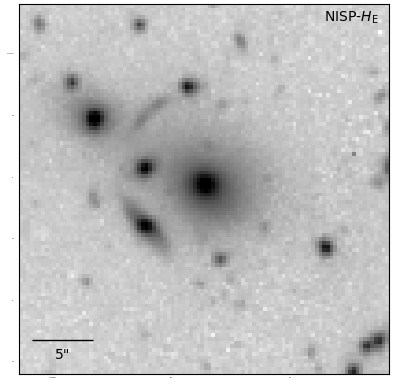}
\caption{\ang{;;30}$\times$\ang{;;30} cutout centered on the galaxy cluster ACT-CLJ0021.5$-$4902 at $z_{\rm spec}=0.688$, associated with the eROSITA DR1 source 1eRASSJ002132.6$-$490216, as seen in VIS \IE\ (top panel) and NISP \HE\ (bottom panel). We can clearly identify a strong lensing arc that was previously unknown.  
}
\label{fig:source2ACTeRosita}
\end{figure}

\begin{figure}[htbp!]
\centering
\includegraphics[angle=0,width=1.01\hsize]{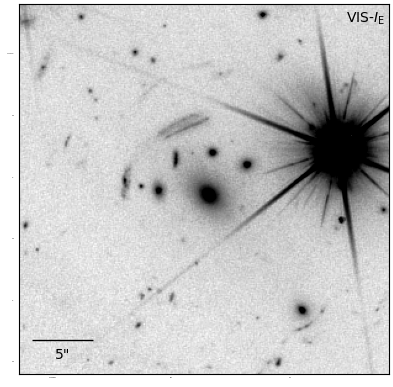}
\includegraphics[angle=0,width=1.01\hsize]{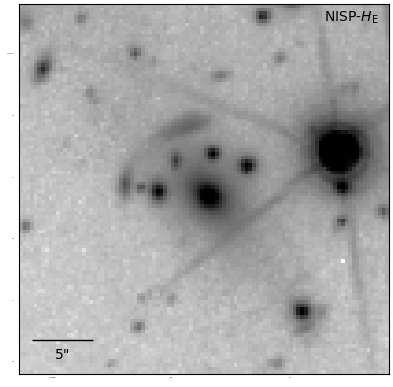}
\caption{\ang{;;30}$\times$\ang{;;30} cutout centered on the galaxy cluster 
ACT-CLJ0023.7$-$4923 with $z_{\rm phot}=0.709 \pm 0.021$, associated with the eROSITA DR1 source 1eRASSJ002345.0$-$492354, as seen in VIS \IE\ (top panel) and NISP \HE\ (bottom panel). Multiple lensing features, which are already known, are visible in the image. The star on the right of the image is $Gaia$ 4973928409193111936 of $G_{\rm RP}=14$ mag. 
}
\label{fig:source3ACTeRosita}
\end{figure}

In order to illustrate the unique capabilities of \euclid\ to detect high-redshift ($z>0.6$) galaxy clusters, we have searched for concentrations of galaxies in the VIS and NISP images in the direction of already known high-redshift clusters. To do so, we have focused on the increasingly large numbers of clusters that are blindly
detected with surveys in the millimeter spectral range via the Sunyaev--Zeldovich effect \citep[SZ;][]{sunyaev1972,sunyaev1972b,sunyaev1980, planck16_cosmo,planck16_reionization,bleem2024} and in the X-ray range \cite[e.g.][]{bulbul2024}. More specifically, we used ground-based telescopes such as the Atacama Cosmology Telescope \citep[ACT, e.g.][]{hilton2021} and the South Pole Telescope \citep[SPT, e.g.][]{bleem2015} which allow us to detect clusters out to $z\sim 1.9$. We also make use of the eROSITA-DE DR1 point and extended sources \citep{merloni2024}. As shown in \citet{bulbul2024}, the first release of eROSITA data is already able to identify groups and clusters at high redshift, out to $z\sim 1.3$. 
 
 The ERO field of A2764 is located in both the ACT and SPT footprints. We searched for the presence of clusters from these two surveys in the ERO observation. We found that the A2764 ERO field contains a few background galaxy clusters previously detected in SZ with masses $M_{500}\sim 2 \times 10^{14} \, M_{\odot}$, located in the redshift range $z\sim 0.6$--0.7, which is higher than the target cluster. 
Three of them are ACT clusters \citep[DR5]{hilton2021}: (1) ACT-CLJ0022.5$-$4843 is an ACT-only cluster at $z_{\rm phot}=0.855 \pm 0.015$ with an estimated mass of $M_{500}=1.71^{+0.37}_{-0.30} \times 10^{14} \, M_{\odot}$; (2) ACT-CLJ0021.5$-$4902 at $z_{\rm spec}=0.688$ and $M_{500}=2.35^{+0.45}_{-0.38} \times 10^{14} \, M_{\odot}$ is also identified as SPT cluster SPT-CLJ0021.5$-$4902 \citep[see][]{bleem2015,everett2020} and it is associated with an eROSITA point-source (see below); and (3) ACT-CLJ0023.7$-$4923 with photometric redshift $z_{\rm phot}=0.709 \pm 0.021$ and mass $M_{500}=1.88^{+0.41}_{-0.33} \times 10^{14} \, M_{\odot}$. These three clusters are clearly identified in the VIS and NISP images as concentrations of galaxies. Moreover, we note that both ACT-CLJ0023.7$-$4923 and ACT-CLJ0021.5$-$4902 show strong gravitational lensing features observed both in the VIS and NISP images (see Figs.~\ref{fig:source2ACTeRosita} and \ref{fig:source3ACTeRosita}). To our knowledge, only the strong lensing in ACT-CLJ0023.7$-$4923 was previously known \citep{diehl2017}, thus the strong lensing observed in ACT-CLJ0021.5$-$4902 is a first detection for this cluster. Finally, a faint SPT source, SPT-SJ002330$-$4947.3, detected at 150~GHz \citep{everett2020} is located near the noisy edge of the A2764 \euclid\ ERO field. The quality of the images does not allow us to identify the possible visible or NIR counterparts.   

More than 40 point and extended sources from the eROSITA DR1 catalog \citep{merloni2024} are found in the A2764 ERO field. Obviously, the cluster A2764 itself is among the eROSITA DR1 sources in the field. It appears as the two extended sources 1eRASSJ002035.1$-$491245 and 1eRASSJ002032.1$-$491432. The two ACT galaxy clusters \citep{hilton2021} are matched with eROSITA DR1 point-sources: ACT-CLJ0021.5$-$4902 at $z_{\rm spec}=0.688$ is associated with 1eRASSJ002132.6$-$490216 (IDCluster=758, EXT=0) and ACT-CLJ0023.7$-$4923 at $z_{\rm phot}=0.709 \pm 0.021$ is associated with 1eRASSJ002345.0$-$492354 (IDCluster=273 EXT=0). Only one source labelled as extended in the eROSITA DR1 catalog is found in the A2764 ERO field, and would qualify as a galaxy cluster: 1eRASSJ002223.6$-$492746 (with IDCluster=183, EXT=19.3 and EXT\_LIKE=3.07). The VIS and NISP images show concentrations of galaxies at the associated location, and the distribution of photometric redshifts indicates a peak at $z_{\rm phot} \sim 0.8$. We also checked for counterparts in the Massive and Distant Clusters of WISE Survey (MaDCoWs) catalogs \citep{gonzalez19} and didn't find any match in the ERO A2764 field.

Another remarkable eROSITA DR1 point source is 1eRASSJ002212.5$-$492811 (IDCluster=47 EXT=0). It is less than $2\arcmin$ away from the aforementioned source 1eRASSJ002223.6$-$492746 and about 40 VIS and NISP galaxies are located in its vicinity, half a dozen with a photometric redshift peaking at $z_{\rm phot} \sim 0.8$ and about 30 at $z_{\rm phot} \sim 1.4$. Although the association of \Euclid\ galaxies and this eROSITA source is not yet confirmed, it illustrates the power of the \Euclid\ survey in combination with other wavelength datasets to efficiently provide robust galaxy cluster candidates.

This simple search for optical and near-IR counterparts to already known SZ clusters at $z>0.6$ with $M_{500} > 10^{14} \, M_{\odot}$ in this ERO program data demonstrates that \Euclid\ VIS and NISP instruments are particularly efficient at detecting clusters at $z>0.6$. Matching with eROSITA extended and point-sources also opens the way for the detection of new clusters at potentially even higher redshifts. Further work is needed to confirm associations between optical/NIR \Euclid\ galaxies with millimeter or X-ray-detected clusters. We have also yet to run dedicated cluster-finding algorithms to find galaxy clusters and proto-clusters in the \Euclid\ data. Nevertheless, this first analysis demonstrates that \Euclid\ data will be a treasure trove for cluster studies.

\section{Intracluster light}
\label{sec:icl}
\begin{figure*}[htbp!]
\centering
\includegraphics[angle=0,width=1.0\hsize]{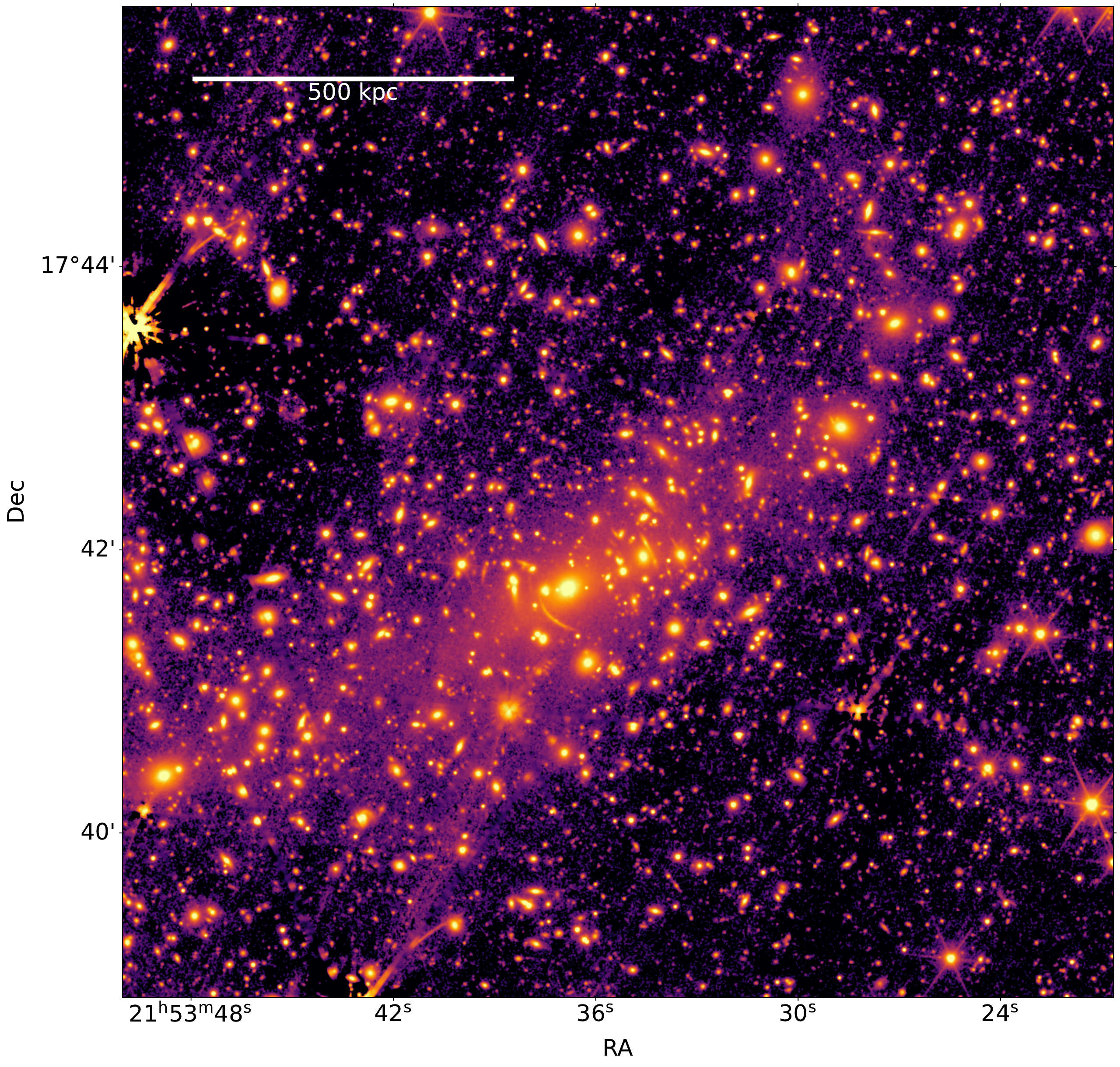}
\caption{\HE-band image showing the extent of ICL in A2390. Stars in the field have been subtracted using the PSF model of \citet{EROData} and the diffraction spikes have been masked by hand. The lognormal color scale is chosen to highlight the diffuse ICL in this cluster.
}
\label{fig:ICL_A2390}
\end{figure*}

Intracluster light (ICL) is diffuse light seen in galaxy groups and clusters that comes from stars that are bound to the gravitational potential well of the cluster rather than any particular galaxy \citep[see][for reviews]{Contini2021, Montes2022}.  It is a prominent feature observed in clusters \citep[e.g. ][]{Feldmeier2004, Kluge2020, Golden-Marx2023}, and the origin and assembly history of the ICL serves as a tool to characterize the assembly history of the brightest central galaxy and the satellite galaxy population \citep[e.g., ][]{Murante2007}. In addition, the ICL can be used as a luminous tracer of the dark matter distribution within the cluster \citep{MT19, Alonso-Asensio2020, Deason2021, Gonzalez2021, Yoo2022, ContrerasSantos2024}. 

Although the ICL appears to be a powerful tool for studying galaxy clusters and cosmology, our understanding of this component has been hampered by observational limitations. For example, ground-based observations have been challenging due to the extreme faintness of this emission. HST has provided ultra-deep images of galaxy clusters, but it has a clear limitation: its small field of view. The high resolution, space-based images of \Euclid\ can characterize the ICL beyond the typical HST limit of 100--200 kpc and will help improve our understanding of the formation of this diffuse light and the assembly of clusters.  In this sense, \Euclid\ can finally provide the wide field of view needed to probe the outer regions of clusters. Using observations from this ERO program, we will probe the ICL of A2390, A2764 and the other high redshift clusters detected in the field of A2764 out to distances of several hundred kpc. 

Figure \ref{fig:ICL_A2390} shows the extent and distribution of the ICL surrounding A2390 in the $\HE$ band. We used the outer PSF model of \citet{EROData} to subtract the PSF contribution of several bright stars in the field. To highlight the low surface-brightness ICL, we masked the remaining diffraction spikes by hand and interpolated over the masked pixels using a Gaussian kernel with $\sigma=0\farcs6$. The ICL stretches over approximately 1\,Mpc, from southwest to northwest, following the distribution of the brightest members of the cluster.  The $1\sigma$ image depth is $\mu(\HE)=28.88$\,mag\,arcsec$^{-2}$ measured over a $10\arcsec\times10\arcsec$ area; see \citet[][]{EROData} for details. At this depth, the main sources that limit the reliability of the ICL measurement are Galactic cirri and noise from the subtraction of the NIR persistence. The ERO data on the Perseus cluster reached a similar depth as these ERO data on A2390 and A2764 \citep{EROData}. The ICL of the Perseus cluster was reliably measured out to a semi-major axis distance of 600\,kpc reaching a $1\,\sigma$ depth of $\mu(\IE)=29.4$\,mag\,arcsec$^{-2}$, $\mu(\YE)=28.4$\,mag\,arcsec$^{-2}$, and $\mu(\JE)=\mu(\HE)=28.7$\,mag\,arcsec$^{-2}$ \citep{EROPerseusICL} in similar cirri conditions and significantly more persistence (due to larger galaxies in the field of view). Hence, we expect to reach similar depths in these images. 
 

\section{Summary}
\label{sec:summary}

The ERO program presents an opportunity to highlight the capabilities and performance of the \Euclid\ spacecraft, offering a glimpse into the nominal survey observations. In this paper, we utilize data acquired from the ERO11 program, which focused on two lensing clusters, A2390 and A2764. The nature and depth of these data enable exploration of various scientific cases related to galaxy formation and evolution, weak and strong lensing phenomena, as well as the physics of galaxy clusters. We have produced a photometric catalog in \IE, \YE, \JE, and $\HE$ bands for a total of 501\,994 sources using different aperture sizes. We also performed multi-band photometry based on VIS detection, which include supporting ground-based observations. These catalogs are made publicly available along with this paper.  

The combination of NISP and VIS imaging is well-suited for identifying dropout sources that are visible only in the NISP images and remain undetected in the optical $\IE$ filter. This is particularly relevant for detecting the Lyman break of high-redshift sources at $z>6$, since the survey design naturally produces roughly 3 magnitudes breaks. Combining photometric criteria with SED-fitting with {\tt Eazy}, we have identified a total of 30 LBG candidates. The compactness of these sources, which remain unresolved in the NISP images, presents a challenge in discriminating between galaxies and quasars. Furthermore, despite exhibiting an apparently pronounced Lyman break, certain rare brown dwarfs can display colors resembling those of our targeted sources. Nevertheless, considering that \euclid\ possesses the unique capability to efficiently identify such bright and rare sources on the sky at $z>6$, it becomes clear that this will facilitate extensive spectroscopic follow-up campaigns from both ground- and space-based observatories, enabling the confirmation of their redshifts and the study of their physical characteristics.

These cluster fields also offer the possibility to perform a joint SL+WL analysis to model the mass distribution of these structures. We have combined an already existing SL model of A2390 with our own WL analysis to extend constraints to larger radius and better estimate the virial mass of the cluster. Using the {\tt WSLAP+} code, we show how WL improves the mass profile reconstruction of the cluster compared to SL only. 

We have also illustrate \euclid's potential for detecting high-redshift galaxy clusters by using these ERO data to search for optical and NIR counterparts of known clusters. In particular, we have identified three clusters in the A2764 field, which were previously detected by their X-ray emission (eROSITA) or by the SZ effect (ACT and SPT). Two of the three clusters, which are located at \mbox{$z=0.6$--0.9} show clear gravitational lensing features. This early work already demonstrates the ability of multi-wavelength efforts to discover high-redshift clusters and to calibrate, in the near-future, automatic identification algorithms.

Finally, cluster fields are ideal laboratories to study the ICL, which helps constrain the assembly history of cluster galaxies and to indirectly map the dark matter distribution. We illustrate the large extent of the ICL in A2390 visible in the \Euclid\ images, which go far beyond the limited field of view of HST. 

Overall, this paper underscores the large array of scientific studies that will be carried out in blank (but also lensed) fields with \euclid. The depth, angular resolution, and wide field of view are clearly a unique combination that will push many frontiers in the landscape of \euclid's legacy science.

\begin{acknowledgements} 
HA is supported by the French Centre National d'Etudes Spatial (CNES). The Cosmic Dawn Center (DAWN) is funded by the Danish National Research Foundation (DNRF140). This work has made use of the \texttt{CANDIDE} Cluster at the \textit{Institut d'Astrophysique de Paris} (IAP), made possible by grants from the PNCG and the region of Île de France through the program DIM-ACAV+, and the Cosmic Dawn Center and maintained by S. Rouberol. HM is supported by JSPS KAKENHI Grant Number JP23H00108, and JSPS Core-to-Core Program Grant Numbers JPJSCCA20200002 and JPJSCCA20210003. MJ is supported by the United Kingdom Research and Innovation (UKRI) Future Leaders Fellowship `Using Cosmic Beasts to uncover the Nature of Dark Matter' (grants number MR/S017216/1 \& MR/X006069/1).TS acknowledges  support provided by the Austrian Research Promotion Agency (FFG) and the Federal Ministry of the Republic of Austria for Climate Action, Environment, Energy, Mobility, Innovation and Technology (BMK) via the Austrian Space Applications Programme with grant numbers 899537 and 900565, as
well as the German Research Foundation (DFG) under grant 415537506, and the German Federal Ministry for Economic Affairs and Climate Action (BMWK) provided by DLR under projects no. 50QE2002 and 50QE2302.\\ 
\AckERO\\
\AckEC\\

This work made use of Astropy\footnote{\url{http://www.astropy.org}}: a community-developed core Python package and an ecosystem of tools and resources for astronomy \citep{astropy:2013, astropy:2018, astropy:2022} and Matplotlib \citep{matplotlib07}. This research has made use of Aladin sky atlas, CDS, Strasbourg Astronomical Observatory, France \citep{bonnarel2000}.

\end{acknowledgements}

\bibliography{Euclid,EROplus}

\end{document}